\documentclass[10pt,journal,compsoc]{IEEEtran}

\usepackage[utf8]{inputenc}
\usepackage{graphicx}
\usepackage{amssymb}
\usepackage{array}
\usepackage{multicol}
\usepackage{tabularx}
\usepackage{longtable}
\usepackage{multirow}
\usepackage{amsthm}
\usepackage{xcolor}
\usepackage{subfig}
\usepackage{amsmath}
\usepackage{mathtools}
\usepackage{cuted}
\usepackage[colorinlistoftodos,prependcaption,textsize=footnotesize]{todonotes}
\usepackage{listings}
\usepackage[normalem]{ulem}
\usepackage{soul}
\usepackage{bm}
\usepackage{chngpage}
\usepackage{boldline}
\usepackage{verbatim}
\usepackage{ragged2e}

\theoremstyle{definition}
\newtheorem{definition}{Definition}

\begin{document}

\title{Weak-Key Analysis for BIKE Post-Quantum Key Encapsulation Mechanism}

\author{
Mohammad Reza Nosouhi\IEEEauthorrefmark{1,2},
Syed W. Shah\IEEEauthorrefmark{1,2},
Lei Pan\IEEEauthorrefmark{1},
Yevhen Zolotavkin\IEEEauthorrefmark{1},
Ashish Nanda\IEEEauthorrefmark{1},
Praveen Gauravaram\IEEEauthorrefmark{3},
Robin Doss\IEEEauthorrefmark{2,1},
\IEEEcompsocitemizethanks{\IEEEcompsocthanksitem\IEEEauthorrefmark{1}Centre for Cyber Security Research and Innovation (CSRI), Deakin University, Geelong, VIC 3220, Australia \\
\IEEEcompsocthanksitem\IEEEauthorrefmark{2}Cyber Security Cooperative Research Centre (CSCRC), Australia\hfil\break
\IEEEcompsocthanksitem\IEEEauthorrefmark{3}Tata Consultancy Services (TCS) Ltd., Brisbane, QLD 4000, Australia
}}

\IEEEtitleabstractindextext{%
\justify{
\begin{abstract}
The evolution of quantum computers poses a serious threat to contemporary public-key encryption (PKE) schemes. To address this impending issue, the National Institute of Standards and Technology (NIST) is currently undertaking the Post-Quantum Cryptography (PQC) standardization project intending to evaluate and subsequently standardize the suitable PQC scheme(s). One such attractive approach, called Bit Flipping Key Encapsulation (BIKE), has made to the final round of the competition. Despite having some attractive features, the IND-CCA security of the BIKE depends on the average decoder failure rate (DFR), a higher value of which can facilitate a particular type of side-channel attack. Although the BIKE adopts a Black-Grey-Flip (BGF) decoder that offers a negligible DFR, the effect of weak-keys on the average DFR has not been fully investigated. Therefore, in this paper, we first perform an implementation of the BIKE scheme, and then through extensive experiments show that the weak-keys can be a potential threat to IND-CCA security of the BIKE scheme and thus need attention from the research community prior to standardization. We also propose a key-check algorithm that can potentially supplement the BIKE mechanism and prevent users from generating and adopting weak keys to address this issue.  
\end{abstract}
}
\begin{IEEEkeywords}
BIKE, Black-Gray Flip Decoder, Code-Based Cryptosystems, Post-Quantum Cryptography, Weak-key Analysis 
\end{IEEEkeywords}
}
\maketitle

\section{Introduction and Background}
Key exchange mechanisms (KEMs) play a critical role in the security of the internet and other communication systems. They allow two remote entities to securely agree on a symmetric key without explicit sharing, which can be subsequently used to establish an encrypted session. The currently used KEMs are mostly based on the RSA \cite{rivest1978method} or ECC \cite{koblitz1987elliptic}. RSA and ECC are PKE schemes whose underlying security relies on the difficulty of either integer factorization or discrete logarithm problems. However, these problems can be solved in polynomial time by quantum computation models \cite{shor1994algorithms} with the so-called Cryptographically Relevant Quantum Computers (CRQC) \cite{CRQC}. Thus, it is believed that the current PKE schemes (and KEMs, consequently) will be insecure in the post-quantum era \cite{mosca2018cybersecurity}, \cite{barker2020getting}.\par

To secure KEMs, tremendous research efforts have been made to design quantum-safe PKE schemes. Currently, NIST is also undertaking a standardization project for quantum-safe KEMs. Of numerous approaches, code-based cryptosystems (CBC) are considered a promising alternative to the existing PKE schemes \cite{daniel2015initial}, \cite{fernandez2019pre}. Based on the theory of error-correcting codes, their underlying security relies on that decoding a codeword without the knowledge of an encoding scheme is an $\mathcal{NP}$-complete problem \cite{8012331}. The idea of CBC was incepted by McEliece in 1978 \cite{mceliece1978public} has remained secure against classical and quantum attacks but at the cost of a larger key size. 
To circumvent this issue, Misoczki \emph{et al}.~designed Quasi-Cyclic Moderate Density Parity-Check (QC-MDPC) codes to develop the QC-MDPC variant of the McEliece scheme \cite{misoczki2013mdpc}. This variant has received much attention because of its comparable security with significantly smaller key sizes. Bit Flipping Key Encapsulation (BIKE) \cite{BIKE} mechanism that is submitted to NIST for standardization as a quantum-safe KEM, is built on top of the QC-MDPC variant. Due to its promising security and performance features, BIKE has been selected in the final round of the NIST standardization competition as an alternate candidate \cite{NISTReport}. In addition, BIKE has been recently added to the list of supported KEM schemes for the post-quantum TLS protocol used in the AWS Key Management Service (KMS) offered by Amazon \cite{AWS}. \par

The QC-MDPC variant and BIKE leverage a probabilistic and iterative decoder in their decapsulation modules. The original design of QC-MDPC variant employed the orthodox Bit-Filliping (BF) decoder \cite{gallager1962low} with slight modifications \cite{misoczki2013mdpc}. However, the BF decoder (as a probabilistic and iterative decoder) suffers from higher DFR. Specifically, the decoder suffers from poor decoding performance when the number of iterations is restricted for performance considerations like accelerating the encoder. It fails in the decryption/decapsulation process, which degrades the performance and can also facilitate the side channel/reaction attacks. For example, Guo \emph{et al}.~\cite{guo2016key} introduced an efficient reaction attack for the QC-MDPC variant known as the GJS attack. In a GJS attack, the attacker firstly sends crafted ciphertexts to the victim Alice while observing the reaction of her decoder for every ciphertext (i.e., \emph{successful} or \emph{failure}). Then, utilizing the correlation between the faulty ciphertext patterns and Alice's private key, the attacker can fully recover her private key. Further, Nilsson \emph{et al}.~\cite{nilsson2018error} proposed a novel technique for fast generation of the crafted ciphertexts to improve the efficiency of the GJS attack. This attack can be regarded as a weaker version of Chosen-Ciphertext Attacks (CCA) since the adversary only needs to observe the decoder's reaction without access to the full decryption oracle (i.e., it does not analyze any decrypted plaintext). \par

To tackle these attacks, the Fujisaki-Okamoto (FO) CCA transformation model has been adopted in BIKE \cite{hofheinz2017modular}, \cite{drucker2021applicability}. The FO model uses a ciphertext protection mechanism so that the receiver can check the integrity of a received ciphertext. Thus, the ability of a GJS attacker to craft ciphertexts is limited. Although the FO model significantly mitigates the threat of reaction attacks, the scheme still must deploy a decoder with a negligible DFR to provide the Indistinguishability under Chosen-Ciphertext Attack (IND-CCA) security. Sendrier \emph{et al}.~\cite{sendrier2020low} argued that to provide $\lambda$-bit of IND-CCA security, the \emph{average} DFR (taken over the whole keyspace) must be upper bounded by $2^{-\lambda}$. For this reason, several modifications of the BF decoding algorithms have been proposed to offer negligible DFR, \cite{sendrier2020low, sendrier2019decoding,  drucker2019toolbox, drucker2019constant, drucker2020qc, nilsson2021weighted}. For example, the latest version of BIKE deploys the Black-Grey-Flip (BGF) decoder \cite{drucker2020qc} that is the state-of-the-art variant of the BF algorithm. BGF uses only five iterations for decoding while offering a negligible DFR.\par

However, it is shown that there are some (private) key structures for which the probabilistic decoders show poor performance in terms of DFR \cite{drucker2019constant}, \cite{sendrierexistence}. They are referred to as \emph{weak-keys} since they are potentially at the risk of disclosure through side-channel/reaction attacks (such as GJS attack) \cite{guo2016key}, \cite{nilsson2018error}. Although the number of weak keys is much smaller than the size of the entire keyspace, their effect on the average DFR must be analyzed to ensure they do not endanger the IND-CCA security of the scheme. 
In this regard, Drucker \emph{et \ al.}~\cite{drucker2019constant}, argued that IND-CCA security of BIKE cannot be claimed without first proving (or disproving) the existence of weak keys and formally quantifying their impact on DFR. Subsequently, Sendrier \emph{et al.}~\cite{sendrierexistence} have recently conducted some weak-key analysis for the QC-MDPC-based schemes and showed that the average DFR is not notably affected by the weak-keys (that have been identified so far). Based on their empirical results, they argue that the existence of weak keys are not critical to IND-CCA claims. However, the state-of-the-art BGF decoder has not been investigated in their analysis. For example, Sendrier \emph{et al.}~\cite{sendrierexistence} have considered the previous version of the BIKE scheme that was submitted to the second round of the NIST competition (i.e., BIKE-1). In BIKE-1, the BackFlip decoder \cite{aragon2017bike} was deployed that enables the scheme to provide 128-bit security in 100 iterations (i.e., BackFlip-100). Moreover, the number of iterations has been limited to 20 for time-saving (significantly larger than the BGF decoder) in existing experiments. For compensating the few iterations, their results are compared with 97-bit security, i.e., the estimated security of BackFlip-20.\par

In view of the aforementioned discussion, the existent analysis on weak-keys does not extend towards the latest version of BIKE that adopts the contemporary BGF decoder. Therefore, it is important to investigate the impact of weak-keys on the latest version of BIKE. To the best of our knowledge, the effect of weak-keys on the average DFR of the BGF decoder has not been investigated. Thus, the IND-CCA security claims of the latest version of BIKE remain occluded. Motivated by this, we first implement the BIKE scheme in Matlab. Then, through extensive experiments and based on the model for IND-CCA security presented in \cite{hofheinz2017modular}, we show that the contribution of weak-keys in the average DFR of the BIKE's BGF decoder is greater than the maximum allowed level needed for achieving IND-CCA security. As a result, the negative effect of weak-keys on the average DFR can not be ignored and must be addressed before claiming the IND-CCA security of BIKE.     
To address the weak-keys issue, we also propose a key-check mechanism that can be integrated into the key generation module of the BIKE scheme to ensure the private keys generated by users are not weak. The main contributions of this paper are summarized as follows:
\begin{itemize}
    \item We perform an implementation of the BIKE scheme with the state-of-the-art BGF decoder and provide some technical key points required to implement the BIKE scheme.   
    \item Through extensive experiments and using the formal model for proving the IND-CCA security, we show that the negative effect of weak-keys on average DFR is greater than the maximum allowed level. It may put the IND-CCA security of the BIKE mechanism at risk.
    \item We propose a key-check algorithm that can be integrated into the key generation subroutine of BIKE to ensure that users do not generate and adopt weak (private) keys. 
\end{itemize}
The paper is organized as follows. In Section \ref{Section.2}, we provide the preliminaries required for understanding the working principles of the BIKE scheme. Section \ref{Section.3} presents the structure of weak-keys in the BIKE scheme and an intuitive understanding of their effect on IND-CCA security. In Section \ref{Section.4}, we present the results of our experimental evaluation. Finally, after introducing the key-check mechanism in Section \ref{Section.5}, we make concluding remarks in Section \ref{Section.6}.

\section{Preliminaries}
\label{Section.2}
In this section, we present the basic concepts that will help the readers to understand the other sections of the paper. We first briefly review the QC-MDPC variant of the McEliece scheme (we refer the readers to \cite{zajac2014overview, misoczki2013mdpc} for more information about the McEliece scheme and its QC-MDPC variant). Then, we review the BF decoding algorithm and the state-of-the-art BGF decoder. Finally, we describe the latest version of the BIKE scheme.\par

\subsection{QC-MDPC Codes}
Before we review QC-MDPC codes, we present some key concepts and definitions in the field of error correction codes. Error correction codes are widely used in communication protocols and recording systems to ensure reliable data transmission and storage. Considering a block of $k$ information bits, the error correction code $\mathbb{C}(n,k)$ computes $r=n-k$ redundancy bits (based on some encoding equations) and creates an $n$-bit block of data (called a codeword) consisted of $k$ information bit and $r$ redundancy bits. The codeword is subsequently sent to the relevant destination, which exploits the redundant bits (based on some decoding rules) for detecting and correcting any errors in the received message and successfully retrieves the actual information.\par

\begin{definition} [Linear Block Code \cite{costello2007channel}] $\mathbb{C}(n,k)$ is a linear error correction code if the modulo-2 sum (i.e., XOR binary operation) of any two or multiple codewords is a valid codeword. 
\end{definition}

\begin{definition} [Hamming weight \cite{bonello2010low}] The Hamming weight of a codeword is defined as the number of non-zero bits in the codeword. 
\end{definition}

\begin{definition} [Generator Matrix \cite{costello2007channel}] The linear block code $\mathbb{C}(n,k)$ has a generator matrix $\textbf{G}\in \mathbb{F}_{2}^{k\times n}$ which defines the one-to-one mapping between the $k$-bit message block $\textbf{m}\in \mathbb{F}_{2}^{k}$ and the corresponding $n$-bit codeword $\textbf{c} \in \mathbb{F}_{2}^{n}$, i.e., $\textbf{c}_{1\times n} = \textbf{m}_{1\times k}.\textbf{G}_{k\times n}$.
\end{definition}

Thus, $\textbf{G}$ is used by the encoder to generate the distinct codeword $\textbf{c}$ associated with the message block $\textbf{m}$. Note that the number of valid codewords for $\mathbb{C}(n,k)$ is $2^k$ which can be much smaller than $2^n$ (since $n>k$) , i.e., every binary vector over $\mathbb{F}_{2}^{n}$ is not necessarily a valid codeword of $\mathbb{C}$. Thus, $\mathbb{C}$ can be considered as a $k$-dimensional subset of $\mathbb{F}_{2}^{n}$. 

\begin{definition} [Systematic Code \cite{ fang2015survey}] $\mathbb{C}(n,k)$ is called a systematic code if its generator matrix is written in the form of $\textbf{G} = [\textbf{I}_k |\textbf{A}_{k\times r}]$ in which $\textbf{I}_k$ is a $k\times k$ identity matrix and $\textbf{A}$ is a $k\times r$ coefficient matrix.
\end{definition}

If $\mathbb{C}$ is systematic, in each $n$-bit codeword $\textbf{c}$, the first $k$ bits are equal to the corresponding message block $\textbf{m}$, and the rest of the block is the $r=n-k$ parity-check (redundant) bits.

\begin{definition} [Parity-check Matrix \cite{costello2007channel}] The parity-check matrix $\textbf{H}\in \mathbb{F}_{2}^{r\times n}$ of a linear code $\mathbb{C}(n,k)$ is an $r\times n$ matrix that is orthogonal to all the codewords of $\mathbb{C}(n,k)$, i.e., $\textbf{c}$ is a valid codeword of $\mathbb{C}(n,k)$ if and only if $\textbf{c}.\textbf{H}^T = \textbf{0}$, where $T$ denotes the matrix transpose operation.
\end{definition}

If $\textbf{G}$ is written in the systematic form (i.e., $\textbf{G} = [\textbf{I}_k |\textbf{A}_{k\times r}]$), it is shown that $\textbf{H}$ can be computed through $\textbf{H} = [\textbf{A}^T|\textbf{I}_r]$. The decoder of $\mathbb{C}$ uses $\textbf{H}$ to decode the received vector.

\begin{definition} [Syndrome of a received vector \cite{costello2007channel}] Consider $\textbf{x}=(\textbf{c}\oplus \textbf{e})\in \mathbb{F}_{2}^{n}$  as a vector received by the decoder, where $\textbf{e}\in \mathbb{F}_{2}^{n}$ is the error vector with the maximum Hamming weight $t$ that represents the flipped bits of $\textbf{c}$ due to the noisy channel. The syndrome $\textbf{S}\in \mathbb{F}_{2}^{r}$ of $\textbf{x}$ is computed as $\textbf{S} =\textbf{x}.\textbf{H}^T$.
\end{definition}

For the syndrome vector $\textbf{S}$, we have
\begin{equation*}
    \textbf{S} =\textbf{x}.\textbf{H}^T = (\textbf{c}\oplus \textbf{e}).\textbf{H}^T = \textbf{c}.\textbf{H}^T \oplus \textbf{e}.\textbf{H}^T =\textbf{e}.\textbf{H}^T,
\end{equation*}
because $\textbf{c}.\textbf{H}^T=\textbf{0}$. Thus, once $\textbf{x}$ is received by the decoder, its syndrome $\textbf{S}$ is firstly computed through $\textbf{S} =\textbf{x}.\textbf{H}^T$. Then, the decoder needs to obtain $\textbf{e}$ by solving $\textbf{S} =\textbf{e}.\textbf{H}^T$ which is then used to compute the sent codeword using $\textbf{c}=\textbf{x}\oplus \textbf{e}$. Finally, the message block $\textbf{m}$ associated with $\textbf{c}$ is returned as the decoded vector. 

\begin{definition} [Syndrome Decoding (SD) Problem \cite{berlekamp1978inherent}] Given the parity-check matrix $\textbf{H}\in \mathbb{F}_{2}^{r\times n}$ and the syndrome vector $\textbf{S}\in \mathbb{F}_{2}^{r}$, the SD problem searches for a vector $\textbf{e}\in \mathbb{F}_{2}^{n}$ with the Hamming weight $\leq t$ such that $\textbf{S}=\textbf{e}.\textbf{H}^T$.
\end{definition}

The SD problem was proved to be $\mathcal{NP}$-complete if the parity-check matrix $\textbf{H}$ is random \cite{berlekamp1978inherent}. This establishes the essential security feature required by code-based cryptosystems to be quantum-resistant. This is because quantum computation models are considered to be unable to efficiently solve $\mathcal{NP}$-complete problems \cite{aaronson2008limits}.

\begin{definition} [Quasi-cyclic (QC) code \cite{baldi2014qc}] The binary linear code $\mathbb{C}(n,k)$ is QC if there exists an integer $n_0<n$ such that every cyclic shift of a codeword $\textbf{c}\in \mathbb{C}$ by $n_0$ bits results in another valid codeword of $\mathbb{C}$.
\end{definition}

In a systematic QC code, each codeword $\textbf{c}$ consists of $p$ blocks of $n_0$ bits, i.e., $n= n_0p$. Thus, every block includes $k_0=k/p$ information bits and $r_0=n_0-k_0$ parity bits. In a QC code $\mathbb{C}$ with $r_0=1$, we have
\begin{equation*}
    r=(n-k)=(n_0-k_0)p=r_0p=p
\end{equation*}
In this case, it is shown that the parity-check matrix $\textbf{H}$ of $\mathbb{C}$ is composed of $n_0$ circulant blocks of size $p\times p$ (or $r\times r$, equivalently) \cite{baldi2014qc} which is written as 
\begin{equation} \label{eq-1}
    \textbf{H} = [\textbf{H}_0 \ \textbf{H}_1 \ldots \textbf{H}_{n_{0}-1}],
\end{equation}
where each circulant block $\textbf{H}_i$ has the following format: 
\begin{equation} \label{eq-2}
    \textbf{H}_i=\begin{bmatrix}
     h_0^{(i)}& h_1^{(i)} & h_2^{(i)} & \ldots & h_{r-1}^{(i)}\\
     h_{r-1}^{(i)}& h_0^{(i)} & h_1^{(i)} & \ldots & h_{r-2}^{(i)}\\
     . & . & . & . & .\\
     . & . & . & . & .\\
     . & . & . & . & .\\
     h_1^{(i)}& h_2^{(i)} & h_3^{(i)} & \ldots & h_0^{(i)}
    \end{bmatrix}
\end{equation}
Note that $\textbf{H}_i$ can be described by its first row only, i.e., the other $r-1$ rows are obtained by cyclic shifts of the first row. It is also shown that the generator matrix $\textbf{G}$ of the above QC code $\mathbb{C}$ can be written as
\begin{equation} \label{eq-3}
    \textbf{G}=[\textbf{I}_{k}|\textbf{Q}_{k\times r}]=\begin{bmatrix}
     & | & (\textbf{H}_{n_{0}-1}^{-1}.\textbf{H}_0)^T\\
     & | & (\textbf{H}_{n_{0}-1}^{-1}.\textbf{H}_1)^T\\
     & | & (\textbf{H}_{n_{0}-1}^{-1}.\textbf{H}_2)^T\\
     \textbf{I}_{k}& | & .\\
     & | & .\\
     & | & .\\
     & | & (\textbf{H}_{n_{0}-1}^{-1}.\textbf{H}_{n_{0}-2})^T\\
    \end{bmatrix}
\end{equation}
The above format can be proved using the fact that $\textbf{H}\textbf{G}^T=\textbf{0}$ and by performing some linear algebra operations on it.

\begin{definition} [QC-MDPC codes \cite{misoczki2013mdpc}] An $(n,r,w)$-QC-MDPC code is a QC code of length $n=n_0r$ and dimension $k=n-r=k_{0}r$ whose parity-check matrix has a constant row weight of $w=O(\sqrt{n})$.
\end{definition}

\noindent Note that here, we only consider those QC-MDPC codes in which $r_0=1$, i.e., $r=p$. \par
The most important characteristics of QC-MDPC codes is that the circulant blocks in the parity-check matrix can be described by their first row only ($r$ bits only). Thus, to construct $\textbf{H}$, one needs only the first row of the $n_0$ circulant blocks. Moreover, the parity-check matrix has a relatively small Hamming weight (i.e., $w<<n$). Therefore, instead of storing $n=n_{0}r$ bits, the positions (indexes) of $w$ non-zero bits can be used to store $\textbf{H}$. These are the key features of QC-MDPC codes that enable them to significantly mitigate the key size issue of the original McEliece scheme. As you will see in the next subsection, private key of the QC-MDPC variant (and BIKE, consequently) is the parity-check matrix of the selected code.\par

In the next subsection, we briefly review the new variant of the McEliece scheme that works based on QC-MDPC codes.

\subsection{The QC-MDPC PKE Scheme}
\begin{figure}[t!]
    \centering
    \includegraphics[width=3.3in, height=1.1in]{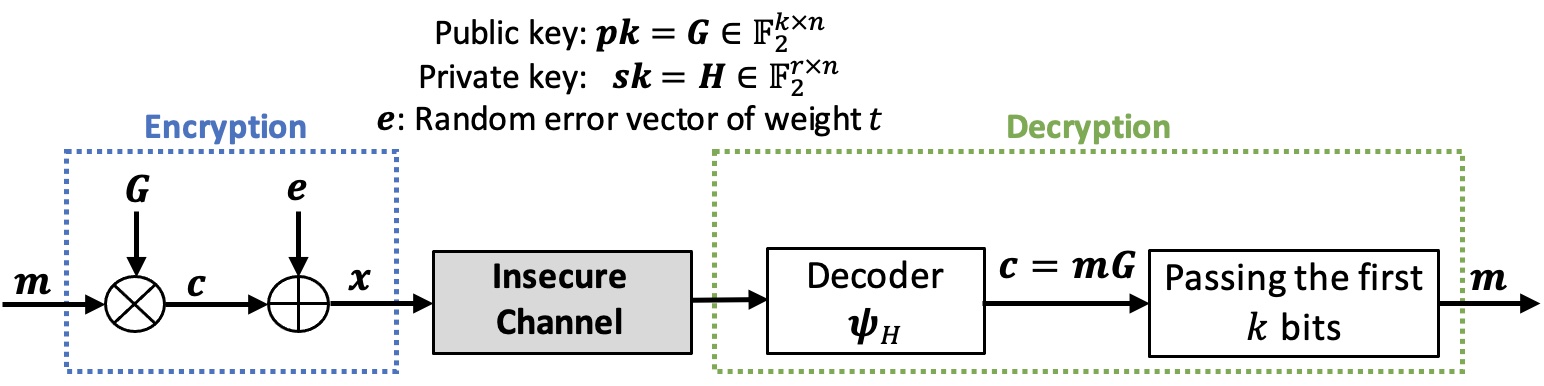}
    \caption{Block diagram of the QC-MDPC variant of the McEliece scheme.}
    \label{fig:1}
\end{figure}

This cryptosystem is a variant of the McEliece code-based PKE scheme. It was proposed by Misoczki et al.~\cite{misoczki2013mdpc} to mitigate the key size problem of the original McEliece scheme. It consisted of three subroutines, i.e., \emph{Key Generation}, \emph{Encryption}, and \emph{Decryption} (see Fig. \ref{fig:1}).

\subsubsection{Key Generation} 
\textbf{Private Key:} In this variant, the parity-check matrix $\textbf{H}$ of the underlying $(r,n,w)$-QC-MDPC code plays the role of private key. It has the format shown in equation (\ref{eq-1}). To generate $\textbf{H}$, $n_0$ circulant blocks of size $r\times r$ must be generated. To do this, for each block $\textbf{H}_i$ ($0\leq i\leq n_0-1$), a random sequence of $r$ bits and Hamming weight $w_i$ is generated such that $\sum_{i=0}^{n_0-1}w_i=w$. This sequence is considered as the first row of $\textbf{H}_i$. Then, the other $r-1$ rows are computed through cyclic shifts of the first row (i.e., $j$ cyclic shifts to generate row $j$, $1\leq j\leq r-1$).\par 

To store the private key, $wlog_2(r)$ bits are needed. This is because each circulant block $\textbf{H}_i$ is represented by its first row only which can be stored using the indexes of its $w_i$ non-zero bits. This is much less than $n^2+k^2+nk$ bits needed to store private key of the original McEliece scheme. \par 

\noindent
\textbf{Public Key:} The generator matrix $\textbf{G}$ of the underlying $(r,n,w)$-QC-MDPC code is the public key of this cryptosystem. It can be computed from the private key using equation (\ref{eq-3}). Since $\textbf{G}$ is quasi-cyclic (similar to the circulant blocks in $\textbf{H}$), it can be represented by its first row only which has $n=k+r$ bits. Note that, the first $k$ bits belong to the identity matrix $\textbf{I}_k$ that do not need to be stored (always have a specific format). Thus, $r$ bits are required to store the public key which shows a significant reduction in the key size compared with $nk$ bits in the original McEliece scheme. Unlike $\textbf{H}$, the first row of $\textbf{G}$ does not necessarily have a small (and fixed) Hamming weight. Thus, the idea of storing the indexes of non-zero bits can not be used for the storage of public key.   

\subsubsection{Encryption}
The encryption of a plaintext message $\textbf{m}\in \mathbb{F}_{2}^{k}$ is performed using the following equation:
\begin{equation*}
    \textbf{x}=\textbf{m}.\textbf{G}\oplus \textbf{e}=\textbf{c}\oplus \textbf{e},
\end{equation*}
where $\textbf{e}\in \mathbb{F}_{2}^{n}$ is a random vector of weight $t$ that is determined based on the error correcting capability of
the corresponding decoder.
\begin{table}[tb!]
	\caption{System parameters of BIKE for different security levels.}
	\label{BIKE_Par}
	\begin{center}
	\resizebox{0.5\textwidth}{!}{%
		\begin{tabular}{|c|c|c|c|c|}
		\hline
			\multirow{2}{*}{\bfseries Parameter} & \multirow{2}{*}{\bfseries Description} & \multicolumn{3}{c|}{\bfseries Value} \\
			\cline{3-5}
			& & Level 1 ($\lambda=128$) & Level 2 ($\lambda=192$)& Level 3 ($\lambda=256$)\\
			\hline
			\multirow{2}{*}{$r$} & Size of circulant blocks in& \multirow{2}{*}{12,323} & \multirow{2}{*}{24,659} & \multirow{2}{*}{40,973}\\
			& the parity-check matrix $\textbf{H}$ & & & \\
            \hline
            \multirow{2}{*}{$w$}& Row weight of the & \multirow{2}{*}{142} & \multirow{2}{*}{206} & \multirow{2}{*}{274}\\
            & parity-check matrix $\textbf{H}$& & & \\
            \hline
            \multirow{2}{*}{$t$} & Hamming weight of & \multirow{2}{*}{134} & \multirow{2}{*}{199} & \multirow{2}{*}{264}\\
            & the error vector & & & \\
            \hline
            \multirow{2}{*}{$l$} & Size of the generated& \multirow{2}{*}{256} & \multirow{2}{*}{256} & \multirow{2}{*}{256}\\
            & symmetric key $\textbf{K}_s$& & & \\
            \hline
		\end{tabular}%
	}
	\end{center}
\end{table}
\subsubsection{Decryption}
The receiver performs the following procedure to decrypt the received ciphertext $\textbf{x}\in \mathbb{F}_{2}^{n}$.
\begin{itemize}
    \item Apply $\textbf{x}$ to the corresponding $t$-error correcting decoder $\psi_{\textbf{H}}$ that leverages the knowledge of $\textbf{H}$ for efficient decoding. The decoder finds the error vector $\textbf{e}$ and returns the corresponding codeword $\textbf{c}=\textbf{m}.\textbf{G}$.   
    \item Return the first $k$ bits of $\textbf{c}$ as the decoded plaintext message $\textbf{m}$ (because $\textbf{G}$ is in the systematic form).
\end{itemize}

Note that the systematic form of $\textbf{G}$ can put the scheme vulnerable to chosen ciphertext and message recovery attacks. The reason is that the first $k$ bits of ciphertext $\textbf{x}=\textbf{m}.\textbf{G}\oplus \textbf{e}$ includes a copy of the plaintext $\textbf{m}$ with some possible flipped bits (since $x_i=m_i\oplus e_i$ for $1\leq i\leq k$). In fact, two ciphertexts $\textbf{x}_1$ and $\textbf{x}_2$ are most likely distinguishable if an attacker knows their corresponding plaintexts $\textbf{m}_1$ and $\textbf{m}_2$. In the worst case, if $e_i=0$ for $1 \leq i\leq k$, the ciphertexts are certainly distinguishable. To address this issue, a CCA transformation model can be used (e.g., \cite{kobara2001semantically}) that converts plaintext $\textbf{m}$ to a random vector whose observation brings no useful knowledge for a CCA attacker.\par

Regarding the above mentioned decoder $\psi_{\textbf{H}}$, several decoding algorithms have been proposed so far \cite{gallager1962low, sendrier2020low, sendrier2019decoding, drucker2019toolbox, drucker2019constant, drucker2020qc, nilsson2021weighted}. We refer the readers to \cite{drucker2020qc} and \cite{drucker2019constant} for more information about the most efficient QC-MDPC decoders. 
\begin{figure*}[t!]
    \centering
    \includegraphics[width=0.9\textwidth]{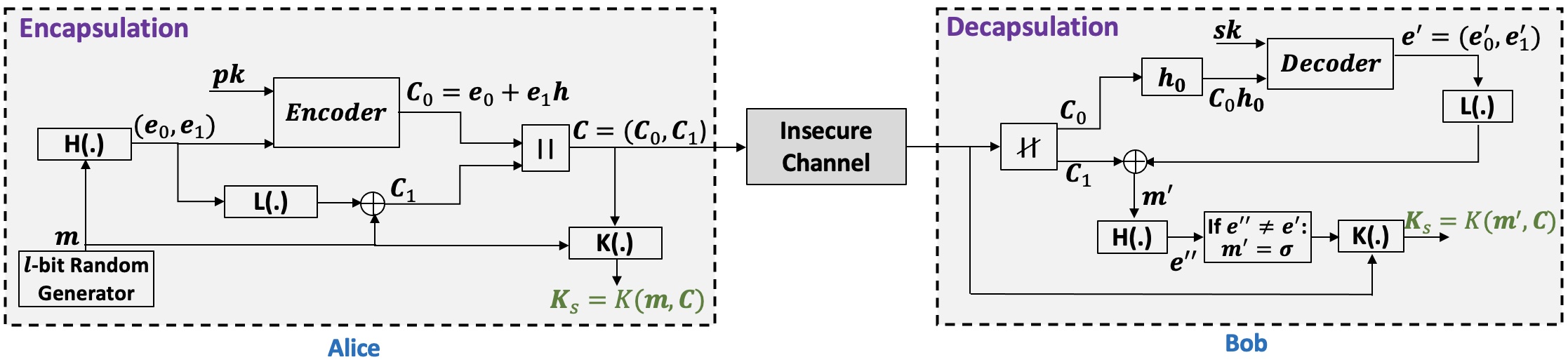}
    \caption{Block diagram of the BIKE scheme. The $\parallel$ and $\nparallel$ symbols represent concatenation and de-concatenation operations, respectively.}
    \label{fig:BIKE}
\end{figure*}
\subsection{The BIKE Scheme}
BIKE \cite{BIKE} is a code-based KEM scheme that leverages the QC-MDPC PKE scheme for encryption/decryption. It has been recently qualified for the final round of the NIST standardization project as an alternate candidate. In previous rounds of the NIST competition, BIKE was submitted in the form of three different versions (BIKE-1, BIKE-2, and BIKE-3) each one satisfied the needs of a specific group of cryptographic applications (e.g., bandwidth, latency, security, etc.). However, in the final round, following the recommendation of NIST, it was submitted in the form of a single version that relies heavily on BIKE-2. The final version suggests three sets of system parameters to satisfy the three different security levels defined by NIST, i.e., level 1 (128-bit security), level 2 (192-bit security), and level 3 (256-bit security) (see Table \ref{BIKE_Par}). To address the IND-CCA security issues exist in the QC-MDPC variant, the Fujisaki-Okamoto (FO) CCA transformation model has been integrated into the BIKE scheme \cite{hofheinz2017modular}, \cite{drucker2021applicability} (see Fig. \ref{fig:BIKE}). In addition, in the final version of BIKE, the state-of-the-art BGF decoder \cite{drucker2020qc} has been deployed in the decapsulation subroutine that provides negligible DFR in five iterations.\par

BIKE includes three subroutines, namely, key generation, encapsulation, and decapsulation (see Fig \ref{fig:BIKE}). The procedure is started by Bob who wants to establish an encrypted session with Alice. They need to securely share a symmetric key to start their encrypted session. To do this, Bob firstly generates his public and private keys by running the key generation subroutine. Then, he sends his public key to Alice who uses it to generate the ciphertext $\textbf{C}=(\textbf{C}_0, \textbf{C}_1)$ and the symmetric key $\textbf{K}_s$ (using the encapsulation subroutine). The first part of $\textbf{C}$ (i.e., $\textbf{C}_0$) is the main data encrypted by the underlying ($r,n,w$)-QC-MDPC scheme while the second part (i.e., $\textbf{C}_1$) protects it against malicious manipulations. Then, Alice sends $\textbf{C}$ to Bob through an insecure channel. By running the decapsulation subroutine, Bob applies $\textbf{C}_0$ to the corresponding QC-MDPC decoder (i.e., the BGF decoder) to decrypt the data. He also checks the integrity of $\textbf{C}_0$ using $\textbf{C}_1$ to ensure it has not been changed. Finally, Bob could generate the same symmetric key $\textbf{K}_s$ as Alice computed.    

In BIKE, all the circulant matrix blocks (e.g., $\textbf{H}_0$ and $\textbf{H}_1$ of the parity-check matrix) are treated as polynomial rings since it increases the efficiency of the computations required in the key generation, encapsulation, and decapsulation subroutines. In this regard, considering $\textbf{a}$ as the first row of the $r\times r$ circulant matrix $\textbf{A}$, the $r$-bit sequence $\textbf{a}$ can be represented by the polynomial $(a_0+a_1x+a_2x^2+\ldots +a_{r-1}x^{r-1}) \in \mathcal{R}=\mathcal{F}_2[x]/(x^r-1)$ (see \cite{aragon2017bike} for more information). In the following, we briefly review the three subroutines of the BIKE scheme. We refer readers to \cite{BIKE} for more detailed information about BIKE.

\subsubsection{Key Generation} \label{section-key_gen}
\textbf{Private Key:} Since BIKE works based on the QC-MDPC variant, private key is the parity-check matrix of the underlying ($r,n,w$)-QC-MDPC code with $n_0=2$ circulant blocks. To generate the private key, two random polynomial $\textbf{h}_0$ and $\textbf{h}_1$ are generated with the Hamming weight of $|\textbf{h}_0|=|\textbf{h}_1|=w/2$. Then, $\sigma$, a random sequence of $l$ bits is generated. Finally, the private key is set as $\textbf{sk}=(\textbf{h}_0, \textbf{h}_1, \sigma$).\par    

\noindent
\textbf{Public Key:} Set a public key as $\textbf{pk}=\textbf{h}=\textbf{h}_1.\textbf{h}_{0}^{-1}$.

\subsubsection{Encapsulation}\label{section-Encaps}
The encapsulation subroutine takes the public key $\textbf{h}$ as input and generates the ciphertext $\textbf{C}$ and the symmetric key $\textbf{K}_s$. To do this, three hash functions (modelled as random oracles) $\mathcal{H}:\{0,1\}^l\xrightarrow{}$ $\{0,1\}_{|t|}^{2r}$, $\mathcal{L}:\{0,1\}^{2r}\xrightarrow{}$ $\{0,1\}^{l}$, and $\mathcal{K}:\{0,1\}^{2r+l}\xrightarrow{}$ $\{0,1\}^{l}$ are defined and used here. The following procedure is performed in this subroutine.
\begin{itemize}
    \item Randomly select an $l$-bit vector $\textbf{m}$ from the message space $\mathcal{M}=\{0,1\}^l$.
    \item Compute $(\textbf{e}_0, \textbf{e}_1) = \mathcal{H}(\textbf{m})$ where $\textbf{e}_0$ and $\textbf{e}_1$ are error vectors of $r$ bits such that $|\textbf{e}_0|+|\textbf{e}_1|=t$.
    \item Compute $\textbf{C} =(\textbf{C}_0, \textbf{C}_1) = (\textbf{e}_0 + \textbf{e}_1.\textbf{h}, \textbf{m} \oplus \mathcal{L}(\textbf{e}_0, \textbf{e}_1))$ and send it to the recipient.
    \item Compute $\textbf{K}_s = \mathcal{K}(\textbf{m}, \textbf{C})$ as the secret symmetric key.
\end{itemize}

\subsubsection{Decapsulation}
The decapsulation subroutine takes the private key $\textbf{sk}$ and ciphertext $\textbf{C}$ as input and generates the symmetric key $\textbf{K}_s$ as follows.
\begin{itemize}
    \item Decode $\textbf{C}_0$ by computing the syndrome $\textbf{S}=\textbf{C}_0.\textbf{h}_0$ and apply it to the corresponding BGF decoder to obtain the error vectors $\textbf{e}_{0}^{'}$ and $\textbf{e}_{1}^{'}$.
    \item Compute $\textbf{m}^{'}=\textbf{C}_1\oplus \mathcal{L}(\textbf{e}_{0}^{'}, \textbf{e}_{1}^{'})$. If $\mathcal{H}(\textbf{m}^{'})\neq (\textbf{e}_{0}^{'}, \textbf{e}_{1}^{'})$, set $\textbf{m}^{'}=\sigma$.
    \item Compute $\textbf{K}_s = \mathcal{K}(\textbf{m}^{'}, \textbf{C})$
\end{itemize}

The deployed CCA transformation model prevents a CCA attacker (e.g., a GJS attacker) to freely choose any error vector $(\textbf{e}_{0}, \textbf{e}_{1})$ required by the attack procedure and submits the obtained crafted ciphertext to the receiver (i.e., to craft a ciphertext based on a malicious plan). It is because the error vector $(\textbf{e}_{0}, \textbf{e}_{1})$ are computed by the one-way hash function $\mathcal{H}$. If the attacker changes the legitimate error vector $\textbf{e}=(\textbf{e}_{0}, \textbf{e}_{1}^{})$, the integrity check at the receiver will fail (i.e., $\mathcal{H}(\textbf{m}^{'})\neq (\textbf{e}_{0}^{'}, \textbf{e}_{1}^{'})$). Therefore, to feed the ciphertext with a desired error vector $\textbf{e}$, the attacker has to find the corresponding vector $\textbf{m}$ such that $\mathcal{H}(\textbf{m})=\textbf{e}$. This imposes heavy burden to the attacker since many queries must be submitted to the random oracle $\mathcal{H}$ to identify the corresponding vector $\textbf{m}$. We will discuss this problem in the next section.

\section{Weak-Keys in the BIKE Scheme}\label{Section.3}
In this section, we first present a formal definition for the weak-keys that we consider in this work. Then, we discuss the effect of weak-keys on the IND-CCA security of BIKE and review the weak-key structures that have been identified so far. 

\subsection{Definition of Weak-Keys}
In this work, we consider a private key of the BIKE scheme as \emph{weak} if decoding of the ciphertexts generated using its corresponding public key results in a much higher DFR than the average DFR of the decoder. Regardless of performance degradation issues, such weak-keys can offer a significant  advantage to CCA adversary for conducting a reaction attack  (see the next subsection for more details). It is noteworthy that, prior research have indicated the possibilities of recovering weak private keys from the corresponding public key in the QC-MDPC PKE scheme \cite{bardet2016weak}, \cite{9383383}. Specifically, it is shown that there exists  weak private keys whose structure can facilitate an adversary in compromising them by applying some linear algebra techniques such as extended Euclidean algorithm on their corresponding public keys. However, those weak structures are not relevant to IND-CCA security (the adversary does not need to conduct a chosen ciphertext attack to compromise those keys) and thus are not consider in this work. Instead, we assume that recovering private key from the corresponding public key is infeasible in the BIKE scheme. Therefore, the attacker needs to conduct a chosen ciphertext attack to recover the private key by leveraging the decoder's reactions.\par  

\subsubsection{The Negative Effect of Weak-Keys on DFR}
To gain an intuitive insight into understanding the impact of weak-keys upon decoder's performance, we need an elaboration of the following question.\par 

How do the columns of $\textbf{H}$ with a large intersections result in a higher DFR?
To answer this question, we consider $\textbf{H}$ as the parity-check matrix (i.e., private key) in which columns $j$ and $l$ ($j,l\in\{0,1,\ldots,n-1\}, \ j\neq l$) have $m$ non-zero bits at exactly the same positions (i.e., $m$ intersections between the two columns). If $\textbf{H}$ is a normal key (i.e., not a weak key), it is shown that the largest possible value of $m$ is usually small (e.g., 5) as compared with the Hamming weight of each column, i.e., $w/2$ \cite{sendrierexistence}. Now, assume we have private key $\textbf{H}$ in which $m$ (for $j$th and $l$th columns) is much larger than that of normal keys. Also, assume that, $e_j=0$ and $e_l=1$ in the original error vector $\textbf{e}$. In this case, it is intuitive to imagine that the number of unsatisfied parity-check equations (i.e., $upc$) for $j$th and $l$th bits will be highly correlated since they would have similar connections on the corresponding Tanner graph. In other words, both $j$ and $l$ bit nodes are involved in almost the same parity-check equations due to the large intersection between them. Thus, in a decoding iteration, if a specific parity-check equation (that involves bit node $j$ and $l$) is unsatisfied, it will be counted towards both $j$th and $l$th bits. Thus, it is highly likely that the decoder returns $e_j=e_l=1$ due to their (correlated) $upc$s being greater than the set threshold. In fact, a real error at $l$th bit results in a situation that convinces the decoder to incorrectly considers $e_j$ as a set error bit.    
Again, in the next iteration, the same procedure is performed, this time the $j$th bit that was mistakenly considered as a set error bit (i.e., the decoder flipped it to $e_j=1$ in the previous iteration while its real value is 0) results a high value for the correlated $upc_j$ and $upc_l$. Thus, the decoder (again) identifies both of them as set error bits and flips their value. Although this corrects $e_j$, it results in an incorrect value for $e_l$. This process is repeated back and forth for all iterations, making decoder incapable of finding the correct vector $\textbf{e}$ leading to decoding failure.\par  

\subsection{Weak-Keys and IND-CCA Security of BIKE}
As stated earlier, BIKE deploys a probabilistic and iterative decoder with a fixed number of iterations for ensuring that constant-time implementation needed for suppressing any side-channel knowledge available to an adversary \cite{drucker2019constant}. This decoder can fail to successfully decode a ciphertext in the allowed number of iterations. The decoding capability of such probabilistic decoders is generally represented in terms of (average) DFR. Prior research has shown that the higher DFR of a decoder deployed in QC-MDPC-based schemes (such as BIKE) can facilitate the efficient recovery of private key through an attack referred to as GJS (attack) \cite{guo2016key}. In this attack, some specific formats for the error vector $\textbf{e}$ are used to craft a large group of ciphertexts that are submitted to the decryption oracle (this constitutes CCA). Then, utilizing decryption failures, the attacker can recover the private key. As mentioned before, to circumvent this possibility, BIKE mechanism adopts FO CCA transformation model \cite{hofheinz2017modular} that prevents the attacker from cherry-picking the desired error vector $\textbf{e}$ needed for successful conduction of GJS attack ( i.e., in BIKE, $\textbf{e}$ is the output of a one-way hash function, thus forbids attackers from crafting some specifically chosen ciphertexts). However, despite this simple remedy, a lower DFR is still important for ensuring the required level of security (see below for details). \par

For a formal proof of IND-CCA security, we first need to define a $\delta$-correct KEM scheme. Based on the analysis provided in \cite{hofheinz2017modular}, a KEM scheme is  $\delta$-correct if;
\begin{equation}
\label{eq.delta_cor}
\begin{split}
Pr[\emph{Decaps}(\textbf{C}, \textbf{sk})\neq \textbf{K}_s|(\textbf{sk},\textbf{pk}) 
\xleftarrow{} \emph{Key\_Gen}, (\textbf{C}, \textbf{K}_s) 
\xleftarrow{} \\
\emph{Encaps}(\textbf{pk})  
]\leq \delta
\end{split}
\end{equation}
Note that, the term on the left-hand side of the aforementioned inequality is the average DFR (hereinafter denoted as $\overline{DFR}$) taken over the entire key space and all the error vectors. Thus, if $\overline{DFR}\leq \delta$, then the KEM scheme is said to be $\delta$-correct.\par

For a $\delta$-correct KEM scheme, it is shown that the advantage of an IND-CCA adversary $\mathcal{A}$ is upper bounded as,
\begin{equation}\label{eq.IND_CCA}
    Adv_{\text{\tiny $KEM$}}^{\text{\tiny $CCA$}}(\mathcal{A})\leq q.\delta + \beta ,
\end{equation}
where $q$ is the number of queries that $\mathcal{A}$ needs to submit to the random oracle model (i.e., the hash function $\mathcal{H}$) to find the valid vector $\textbf{m}$ that are needed for the desired error vector $\textbf{e}=(\textbf{e}_0,\textbf{e}_1)$ (see Fig.~2). $\beta$ in Eq.~(\ref{eq.IND_CCA}) is a complex term that is related to IND-CPA security and is not relevant to IND-CCA analysis and is thus not considered.\par

Based on the above definitions, a KEM scheme is IND-CCA secure offering $\lambda$-bit security if $\frac{T(\mathcal{A})}{Adv_{KEM}^{CCA}(\mathcal{A})}\geq 2^{\lambda}$ \cite{hofheinz2017modular}, where $T(\mathcal{A})$ is the running time of $\mathcal{A}$ that is approximated as $q.t_q$, with $t_q$ representing the running time of a single query. Since typically $t_q<1$, we have $T(\mathcal{A})<q$, such that $2^{\lambda}\leq \frac{T(\mathcal{A}}{Adv_{KEM}^{CCA}}\leq \frac{q}{q.\delta}$, resulting in $2^{\lambda}\leq \frac{q}{q.\delta}$. Therefore, to provide $\lambda$-bit of IND-CCA security, the KEM scheme must be $\delta$-correct with $\delta\leq 2^{-\lambda}$, or equivalently (using Eq.~(\ref{eq.delta_cor})), 
\begin{equation}\label{eq.DFR}
    \overline{DFR}\leq 2^{-\lambda}.
\end{equation}
However, for QC-MDPC codes used in the BIKE scheme, there exists no known mathematical model for an accurate estimation of $\overline{DFR}$ (taken over the whole key space and error vectors). Instead, $\overline{DFR}$ corresponding to needed security-level is estimated through experiments with limited number of ciphertexts (although sufficiently large) and then applying (linear) extrapolation  (see \cite{sendrier2020low} for more details). As a result, the claimed DFR may necessarily not be same as the actual $\overline{DFR}$. In the worst case, we assume that there is a group of keys for which the value of DFR is high (i.e., the set of weak-keys $\mathcal{K}_w$). If the weak-keys have not been used in the experiments performed for estimating the average DFR, then the actual $\overline{DFR}$ may be larger than the estimated DFR (obtained empirically) such that the condition in Eq.~(\ref{eq.DFR}) is not met. Therefore, the impact of weak keys $\mathcal{K}_w$ on $\overline{DFR}$ must be investigated to estimate the actual value of average DFR and ensure the IND-CCA security of the scheme. To formulate the equation for IND-CCA security in presence of weak-keys, we consider $|\mathcal{K}_w|$ as size of $\mathcal{K}_w$ (i.e., the number of weak-keys), $DFR_w$ as the average DFR taken over $\mathcal{K}_w$, $\mathcal{K}_s$ as the set of other keys (i.e., $\mathcal{K}_w\cup \mathcal{K}_s$ is equal to the whole key space $\mathcal{K}$), and $DFR_s$ as the average DFR taken over $\mathcal{K}_s$. In this case, $\overline{DFR}$ becomes;
\begin{equation}\label{eq.Ave_DFR}
    \overline{DFR}=\eta_s DFR_s + \eta_w DFR_w ,
\end{equation}
where $\eta_s=\frac{|\mathcal{K}_s|}{|\mathcal{K}|}$ and $\eta_w=\frac{|\mathcal{K}_w|}{|\mathcal{K}|}$. By combining Eqs.~(\ref{eq.DFR}) and (\ref{eq.Ave_DFR}), we have, 
\begin{equation}\label{eq.condition}
    \eta_s DFR_s\leq 2^{-\lambda} - \eta_w DFR_w
\end{equation}
From (\ref{eq.condition}), the modified condition for IND-CCA security is obtained as;
\begin{equation}\label{eq.Basic_cond}
    \eta_w DFR_w \leq 2^{-\lambda}
\end{equation}
Therefore, to provide $\lambda$-bit of IND-CCA security, the set of weak-keys $\mathcal{K}_w$ must be negligible enough (compared with $|\mathcal{K}|$) such that  $\eta_w DFR_w < 2^{-\lambda}$ (even if $DFR_w$ is significantly larger than $DFR_s$).

\subsection{Structure of Weak-Keys in BIKE}
The weak-keys for BIKE scheme can be determined by adopting the approach proposed in \cite{guo2016key} (i.e., the GJS attack methodology for recovering private keys). In this attack, the attacker primarily targets $\textbf{H}_0$ (i.e., the first block of private key). If $\textbf{H}_0$ is successfully recovered, then the attacker can easily compute $\textbf{H}_1$ from $\textbf{H}_0$ by performing simple linear algebra operations on $\textbf{G}.\textbf{H}^T=\textbf{0}$. Precisely, using Eq.~(\ref{eq-3}), we can re-write $\textbf{G}.\textbf{H}^T=\textbf{0}$ as $[\textbf{I}_k| \textbf{Q}_{k\times r}].[\textbf{H}_0 \  \textbf{H}_1]^T=\textbf{0}$, which results in $\textbf{H}_{0}^{T}+\textbf{Q}.\textbf{H}_{1}^{T}=\textbf{0}$, since $\textbf{I}_k.\textbf{H}_{0}^{T}=\textbf{H}_{0}^{T}$. Therefore, the attacker can easily obtain $\textbf{H}_1=[\textbf{Q}^{-1}.\textbf{H}_{0}^{T}]^T$ (note that, for $n_0=2$, we have $r=k$).\par

To find $\textbf{H}_0$, the attacker selects the error vector $\textbf{e}$ from a special subset $\boldsymbol{\Psi}_d$ ($d=1,2,\ldots,U$). The parameter $d$ is defined as the distance between two indexes (positions) $i$ and $j$ in the first row of $\textbf{H}_0$ (shown by $\textbf{h}_0$) which is formally defined as follows:
\begin{eqnarray*}
    d(i,j)=min((i-j+r) \ mod \ r, (j-i+r) \ mod \ r) \\
    for \ i,j \in \{0,1,2,\ldots,r-1\}
\end{eqnarray*}
For example, considering $r=10$, the distance between the first and last bits is 1 because $d(0,9)=min((19 \ mod \ 10), (1 \ mod \ 10))=min(9,1)=1$. Based on the above definition, $\boldsymbol{\Psi}_d$ is generated as 
\begin{equation*}
\begin{split}
    \boldsymbol{\Psi}_d=&\{\textbf{e}=(\textbf{e}_0,\textbf{e}_1) \ | \ \textbf{e}_1=\textbf{0}, \exists \ \{p_i\}_{i=1}^{t} \ s.t. \ e_{p_i}=1, \ \ and \\
    & \ \ p_{2i}=(p_{2i-1}+d) \ mod \ r \ for \ i=1,2,\ldots, t/2\}
\end{split}
\end{equation*}
Note that, the second half of the error vector $\textbf{e}$ selected from $\boldsymbol{\Psi}_d$ is an all-$0$ vector, i.e., the Hamming weight of $\textbf{e}_0$ is $t$. Each $p_i$ ($i\in\{1,2,\ldots,t\}$) indicates the position of a non-zero bit in $\textbf{e}_0$.\par  

As demonstrated in prior work research \cite{guo2016key}, when the error vectors are selected from $\boldsymbol{\Psi}_d$, there exists a strong correlation between the decoding failure probability and the existence of distance $d$ between the non-zero bits in $\textbf{h}_0$ (i.e., the first row of first circulate block $\textbf{H}_0$). In other words, if distance $d$ exists between two non-zero bits in $\textbf{h}_0$, then the probability of decoding failure is much smaller in contrast with the case when such a distance $d$ does not exist. Utilizing this important observation, firstly, the attacker (empirically) computes the probability of decoding failure for various distances $d=1,2,\ldots,U$. This is done by submitting many ciphertexts (generated using the error vectors selected from $\boldsymbol{\Psi}_d$) to the decryption oracle and recording the corresponding result of the decryption (i.e., \emph{successful} or \emph{failed}). Then, each value of $d$ is classified into either $\{$\emph{existing}$\}$ or $\{$\emph{not existing}$\}$ classes based on the obtained failure probability, i.e., a specific value of $d$ with a small failure probability is categorized as \emph{existing}, and vice versa. Based on the obtained categorized distances, the distance spectrum of $\textbf{h}_0$ is defined as follows: 
\begin{equation*}
    D(\textbf{h}_0)=\{d: \ 1 \leq d \leq U, \ d \in \{existing\} \ in \ \textbf{h}_0\}.
\end{equation*}
Moreover, since a distance $d$ may appear multiple times in $\textbf{h}_0$, the multiplicity of $d$ in $\textbf{h}_0$ is defined as follows: 
\begin{eqnarray*}
    \mu(d, \textbf{h}_0)=|\{(i,j): \ 0\leq i\leq j\leq r-1, \ h_{0,i}=h_{0,j}=1  \\ and \ d(i,j)=d \}| 
\end{eqnarray*}
Finally, based on the obtained distance spectrum $D(\textbf{h}_0)$ and the multiplicity of every distance $d \in D(\textbf{h}_0)$, the attacker can reconstruct $\textbf{h}_0$. To do this, the attacker assigns the first two non-zero bits of $\textbf{h}_0$ at position 0 and $d_1$, where $d_1$ is the minimum distance in $D(\textbf{h}_0)$. Then, the third non-zero bit is put (iteratively) at a position such that the two distances between the third position and the previous two positions exist in $D(\textbf{h}_0)$. This iterative procedure continues until all the $w/2$ non-zero bits of $\textbf{h}_0$ are placed at their positions. Note that, the attacker needs to perform one or multiple cyclic shifts on the obtained vector to find the actual vector $\textbf{h}_0$. This is because the first non-zero bit was placed at position 0 which is not necessarily the case in $\textbf{h}_0$.\par

The structure of weak-keys in BIKE are determined based on the concepts introduced in the GJS attack (i.e., distance, distance spectrum, and multiplicity of a distance). In this regard, three types of weak-keys have been specified in \cite{sendrierexistence} which are detailed below. 

\subsubsection{Type 1}
\label{t1}
In the first type, considering the polynomial representation of binary vectors, the weak-key $\textbf{h}=(\textbf{h}_0,\textbf{h}_1)$ with $f$ $d$-consecutive non-zero positions and Hamming weight $w$ is defined as \cite{sendrierexistence},   
\begin{equation}\label{eq-Type_1}
    \textbf{h}_i =\phi_d(x^l[(1+x+x^2+\ldots+x^{f-1})+\textbf{h}_{i}^{'}]), \ \ i\in\{0,1\},
\end{equation}
where $d \in\{1,2,,3,\ldots,\lfloor r/2\rfloor\}$ is the distance between non-zero bits of the $f$-bit pattern in the weak-key, $l\in \{0,1,2,\ldots,r-1\}$ determines the beginning position of the $f$-bit pattern, and $\phi_d()$ is a mapping function that replaces $x$ with $x^d$, thus, results in distance $d$ between any two successive $1'$s in $(1+x+x^2+\ldots+x^{f-1})$.\par

Note that, to construct the weak-key in this format, each block $\textbf{h}_{i}; \ i\in\{0,1\}$ (with Hamming weight of $w/2$), is first divided into two sections. The first section is an $f$-bit block in which all the $f$ bits are set to $1$ (i.e., $1+x+x^2+\ldots+x^{f-1}$, in the polynomial form). The second section is $\textbf{h}_{i}^{'}$ which is an $(r-f)$-bit block with the Hamming weight $(w/2-f)$ and randomly chosen nonzero bits. Then, the two sections are concatenated and an $l$-bit cyclic shift is applied on it (using the $x^l$ term). Finally, by applying $\phi_d$, the $f$ 1-consecutive non-zero bits of the first section are mapped to a block of $f$ $d$-consecutive non-zero bits. Note that, in this type of weak-keys, considering the distances $jd$ ($j\in\{1,2,3,\ldots, f-1\}$), a lower bound for the multiplicity metric $\mu$ can be obtained as $\mu(jd,\textbf{h}_i)\geq f-j$ for $d\in \{1,2,,3,\ldots,\lfloor r/2\rfloor\}$. \par

To compute $|\mathcal{K}_w|$, firstly, consider the second section of a weak-key defined using (\ref{eq-Type_1}), i.e., $\textbf{h}_{i}^{'}$. It is a $r-f$-bit vector of Hamming weight $w/2-f$. Thus, we have $r-f \choose w/2-f$ options for $\textbf{h}_{i}^{'}$. For the first part of the weak-key (i.e., the $f$-bit pattern), there is only one option as all the $f$ bits are set to $1$. The entire (concatenated) package is subsequently shifted (cyclically) by $l$ bit, where $l\in \{0,1,2,\ldots,r-1\}$. Thus, we have $r$ different options for the cyclic shifts. Finally, there can be $\lfloor r/2\rfloor$ different mappings for $\phi_d()$, as $d \in\{1,2,,3,\ldots,\lfloor r/2\rfloor\}$. Consequently, for $|\mathcal{K}_w|$, we have \cite{sendrierexistence};
\begin{equation} \label{eq-Num_1}
    |\mathcal{K}_w|(f)\leq 2r\lfloor r/2\rfloor{r-f \choose w/2-f}
\end{equation}
Note that, the $2nd$ term is essential in Eq.~(\ref{eq-Num_1}) because there are two circulant blocks in $\textbf{h}$. Finally, $\eta_w$ for this type is obtained as;
\begin{equation}
    \eta_w (f)= \frac{|\mathcal{K}_w|(f)}{{r \choose w/2}}=\frac{2r\lfloor r/2\rfloor{r-f \choose w/2-f}}{{r \choose w/2}}
    \label{eq:t1}
\end{equation}

\subsubsection{Type 2}
\label{t2}
In the second type of weak-keys identified in \cite{sendrierexistence}, the focus is on having a single distance with a high multiplicity factor. In this type, the weak-key $\textbf{h}=(\textbf{h}_0,\textbf{h}_1)$ with Hamming weight $w$ and parameter $m$ has the multiplicity $\mu(d,\textbf{h}_i)=m$ for a distance $d \in\{1,2,,3,\ldots,\lfloor r/2\rfloor\}$ and $i\in\{0,1\}$.   
If $m=w/2-1$, the number of weak-keys $|\mathcal{K}_w|$ is upper bounded by $2r\lfloor r/2\rfloor$. It is because for $m=w/2-1$, the distance between all the $w/2$ non-zero bits of $\textbf{h}_i$ is $d$. Unlike the first type, the $\textbf{h}_i$ blocks ($i\in\{0,1\}$) do not have a second section $\textbf{h}_{i}^{'}$. Thus, the term ${r-f \choose w/2-f}$ that appeared in Eq. (\ref{eq-Num_1}) is replaced with 1.\par 

However, for $m<w/2-1$, the upper bound for $|\mathcal{K}_w|$ is obtained using a more complicated approach. In this regard, consider the general format $(z_1,o_1,z_2,o_2,\ldots, z_s,o_s)$ for $\textbf{h}_i$ ($i\in \{0,1\}$) that starts with $z_1$ $0'$s followed by $o_1$ $1'$s followed by $z_2$ $0'$s, etc., ($z_i, o_i>0$), i.e., $\sum_{i=1}^{s}o_i=w/2$ and $\sum_{i=1}^{s}z_i=r-w/2$. In this case, it is shown that the upper bound for $|\mathcal{K}_w|$ is \cite{sendrierexistence};  
\begin{equation} 
\begin{split}    |\mathcal{K}_w|\leq 2\lfloor r/2\rfloor \sum_{z_1=1}^{(r-w+m+1)}\sum_{o_1=1}^{(m+1)}(o_1+z_1){w/2-o_1-1\choose s-2}\\
{r-w/2-z_1-1\choose s-2}
    \label{eq:t2}
\end{split}
\end{equation}
where $s$ is the number of $z_i$ and $o_i$ blocks.\par 
\noindent
Eq.~(\ref{eq:t2}) can be proved by applying the \emph{stars and bars} principle \cite{starsbars} on the sets of $\{o_i\}_{i=1}^{s}$ and $\{z_i\}_{i=1}^{s}$ separately . According to this principle, the number of $b$-tuples of positive integers ($x_1,x_2,\ldots,x_b$) whose sum is $N$ (i.e., $\sum_{i=1}^{b}x_i=N$) is;
\begin{equation} \label{eq-star_bar}
    N-1 \choose b-1
\end{equation}
Considering the set $\{o_i\}_{i=1}^{s}$, the value of $o_1$ (the number of bits in the first container) varies from $1$ to $m+1$ and for each value of $o_1$ the principle is applied on the remaining $s-1$ containers, i.e., $b=s-1$ (note that we need to consider the condition $o_1\leq m+1$ to meet $\mu(1,\textbf{h}_i)=m$). Therefore, for every value of $o_1$, we have $b=s-1$ and $N=w/2-o_1$ (because $\sum_{i=1}^{s}o_i=w/2$ and $o_1$ $1$ bits are already allocated to the first container). It results in term $w/2-o_1-1\choose s-2$ in Eq.~(\ref{eq:t2}) (according to (\ref{eq-star_bar})). Similarly, for set $\{z_i\}_{i=1}^{s}$, for every value of $z_1$ ($1\leq z_1\leq r-w+m+1$), we have $b=s-1$ and $N=r-w/2-z_1$ that results in term $r-w/2-z_1-1\choose s-2$ in Eq.~(\ref{eq:t2}). The term $(o_1+z_1)$ is applied to consider the number of different circular shifts that are applicable in each case.\par

\subsubsection{Type 3}
\label{t3}
Unlike the previous types that consider a single block of the parity-check matrix, this type of weak-keys are defined such that the columns of $\textbf{h}_0$ and $\textbf{h}_1$ (in the parity-check matrix) jointly create an ambiguity for the BF-based decoder, resulting in a high DFR. If column $j$ of $\textbf{h}_0$ and column $l$ of $\textbf{h}_1$ both have $m$ non-zero bits at exactly the same positions (i.e., $m$ intersections between the two columns) and $m$ is large, their number of unsatisfied parity-check (i.e., $upc_j$ and $upc_l$) equations (counted during the decoding procedure) will be highly correlated. In this case, if $e_j=1$ or $e_l=1$ in the real error vector $\textbf{e}$, the high level of correlation can prevent the decoder from finding $\textbf{e}$ within the allowed number of iterations. 

The upper bound for $|\mathcal{K}_w|$ in this type of weak keys is obtained as \cite{sendrierexistence}:
\begin{equation} 
    |\mathcal{K}_w|\leq r{w/2\choose m}{r-m\choose w/2-m} 
    \label{eq:t3}
\end{equation}
Firstly, $m$ positions should be chosen from the set of $w/2$ positions of the non-zero bits. Then, once the positions of $m$ non-zero bits are determined, the remaining $w/2-m$ position can be chosen from the remaining $r-m$ available positions. Finally, in every case, $r$ circular shifts of the obtained vector is also a weak-key. It results in the term $r$ in Eq.~(\ref{eq:t3}). Note that, there is no term $2$ in Eq.~(\ref{eq:t3}) (compared with Eqs.~(\ref{eq-Num_1}) and (\ref{eq:t2})) because the second block of $\textbf{h}$ follows the same structure as the first block such that $|\textbf{h}_0\star x^l\textbf{h}_1|=m$ for $l\in\{0,1,2,\ldots, r-1\}$, where $\star$ indicates component-wise product.

\section{Experimental Setup}\label{Section.4}
In this section, we first provide some key technical details about our BIKE implementation. Then, we present the results of our extensive experiments and provide a relevant discussion. 

\subsection{Our Implementation}
\label{impl}
We implemented the key generation, encryption, and decoding modules of the BIKE scheme in MATLAB. We used the BIKE system parameters $(r,w,t)=(12323, 142, 134)$ suggested in the latest technical specification of BIKE \cite{aragon2017bike} for $\lambda=128$ bit security level. The simulations were performed on eight powerful severs equipped with Intel(R) Xeon(R) 2.5GHz CPU (6 processors) and 64GB RAM. We used the extrapolation technique proposed in \cite{sendrier2020low} to estimate DFR of the BGF decoder for the weak-keys. In the following, we provide some technical details on our implementation of key generation, encryption, and decoding procedures.\par

\emph{Key Generation}:
The most difficult challenge in the implementation of key generation module is to perform the polynomial inversion operation (over $\mathbb{F}_2$) which is needed to compute public key from private key (see Eq.~(\ref{eq-3}) and Section \ref{section-key_gen}). Note that, due to the large value of $r$, computing the inverse in the matrix domain is not an efficient approach, and thus can contribute towards computational overhead for our analysis. To have the light-weight inverse operation for our analysis, we adopted the latest extension of the Itoh-Tsujii Inversion (ITI) algorithm \cite{itoh1988fast,drucker2020fast} for polynomial inversion. This algorithm is based on the Fermat's Little Theorem that gives the inverse of polynomial $\textbf{a}=(a_0+a_1x+a_2x^2+\ldots+a_{r-1}x^{r-1})$ as $\textbf{a}^{-1}= \textbf{a}^{2^{r-1}-2}$. The ITI algorithm provides an efficient calculation of $\textbf{a}^{2^i}$  through $i$ cyclic shifts of the $\textbf{a}$'s binary vector. To utilize this approach, the adopted extension of the ITI algorithm uses a novel technique to convert $2^{r-1}-2$ to a series of $2^i$ sub-components that are computed using easy-to-implement cyclic shifts.\par

Based on the mentioned approach, the adopted algorithm needs to perform $\lfloor log(r-1)\rfloor + wt(r-2)-1$ multiplications and $\lfloor log(r-2)\rfloor + wt(r-2)-1$ squaring operations to compute the inverse, where $wt(r-2)$ indicates the Hamming weight of the value of $r-2$ written in the binary format. Thus, it is a scalable algorithm in terms of $r$ and much more efficient than the inverse computation in the matrix domain.\par 

We also perform polynomial multiplications before generating the public key. We developed a technical solution explained in the next paragraph. Finally, for each set of experiments, we saved the generated public and private keys in a file such that the keys can be easily accessed by the encryption and decoding scripts, respectively.\par   

\emph{Encryption}: To compute the ciphertexts, polynomial multiplication is the basic operation performed in the encryption module (see Section \ref{section-Encaps}). We developed the following approach to compute the multiplication of two polynomials $\textbf{a}=(a_0+a_1x+a_2x^2+\ldots+a_{r-1}x^{r-1})$ and $\textbf{b}=(b_0+b_1x+b_2x^2+\ldots+b_{r-1}x^{r-1})$ of degree $r-1$.\par 

Assuming that $\textbf{a}.\textbf{b}=\textbf{c}=(c_0+c_1x+c_2x^2+\ldots+c_{r-1}x^{r-1})$, we computed the binary coefficients of $\textbf{c}$ as $c_0=a_0b_0\oplus a_1b_{r-1}\oplus \ldots \oplus a_{r-1}b_1$, $c_1=a_0b_1\oplus a_1b_0\oplus \ldots \oplus a_{r-1}b_2$, $\ldots$, $c_{r-1}=a_0b_{r-1}\oplus a_1b_{r-2}\oplus \ldots \oplus a_{r-1}b_0$. We observe that
\begin{equation} \label{eq-MUL}
  c_i=\sum_{j=0}^{r-1}a_jb_{(i-j+r) \ mod \ r} \ \ \ mod \ 2 \ \ for \ i\in \{0,1,2,\ldots,r-1\}  
\end{equation}
We implemented Eq.~(\ref{eq-MUL}) to perform the multiplications required for computing ciphertext $\textbf{C}_0$. The same approach is also used to perform the polynomial multiplications in the key generation module.\par

\emph{Decoding}: We implemented a BGF decoder based on Algorithm 1 provided in \cite{aragon2017bike} with 128 bit security level, i.e.,  $\lambda=128$, $NbIter=5$, $\tau=3$. In each iteration, the syndrome vector is firstly updated using the received ciphertext/updated error vector and the private key. Then, the number of unsatisfied parity-check ($upc$) equations for each node bit $i \in \{0,1,\ldots,n-1\}$ is counted and compared with the threshold $T=max(0.0069722.|\textbf{S}| + 13.530, 36)$, where $|\textbf{S}|$ is the Hamming weight of the syndrome vector updated at each iteration. If it is larger than $T$, the relevant bit $i$ in the error vector is flipped. The next iteration is executed using the updated error vector. Finally, after $NbIter=5$ iterations, if the updated syndrome $\textbf{S}$ is equal to $\textbf{e}\textbf{H}^T$, vector $\textbf{e}$ is returned as the recovered error vector; otherwise, the decoder returns \emph{failure}.\par

Note that, in the first iteration of the BGF decoder, two additional steps are  performed that are related to the Black and Gray lists of the bit nodes. These two lists are created and maintained in the first iteration to keep track of those bit flips that are considered uncertain. The Black list includes those bit nodes that have been just flipped (i.e., $upc>T$) while the Gray list maintains the index of those bit nodes for which $upc$ is very close to the threshold such that $upc>T-\tau$. Then, to gain more confidence in the flipped bits, a Black/Gray bit node is flipped if its updated $upc$ is larger than the empirically set threshold $(w/2+1)/2+1$. 
\begin{figure*}[t!]
    \centering
    \includegraphics[width=0.3\textwidth, height=1.3in]{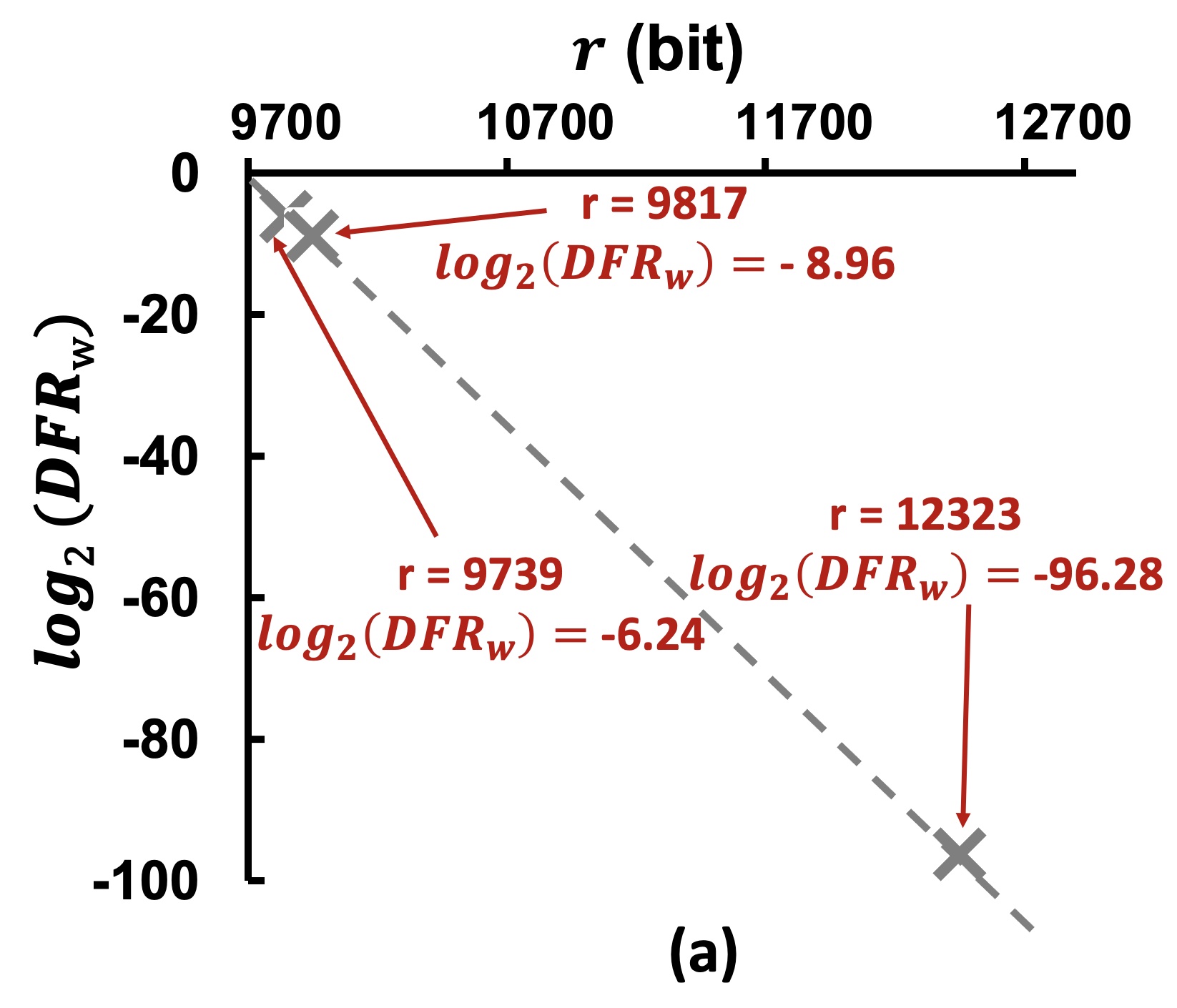}
    \includegraphics[width=0.3\textwidth, height=1.3in]{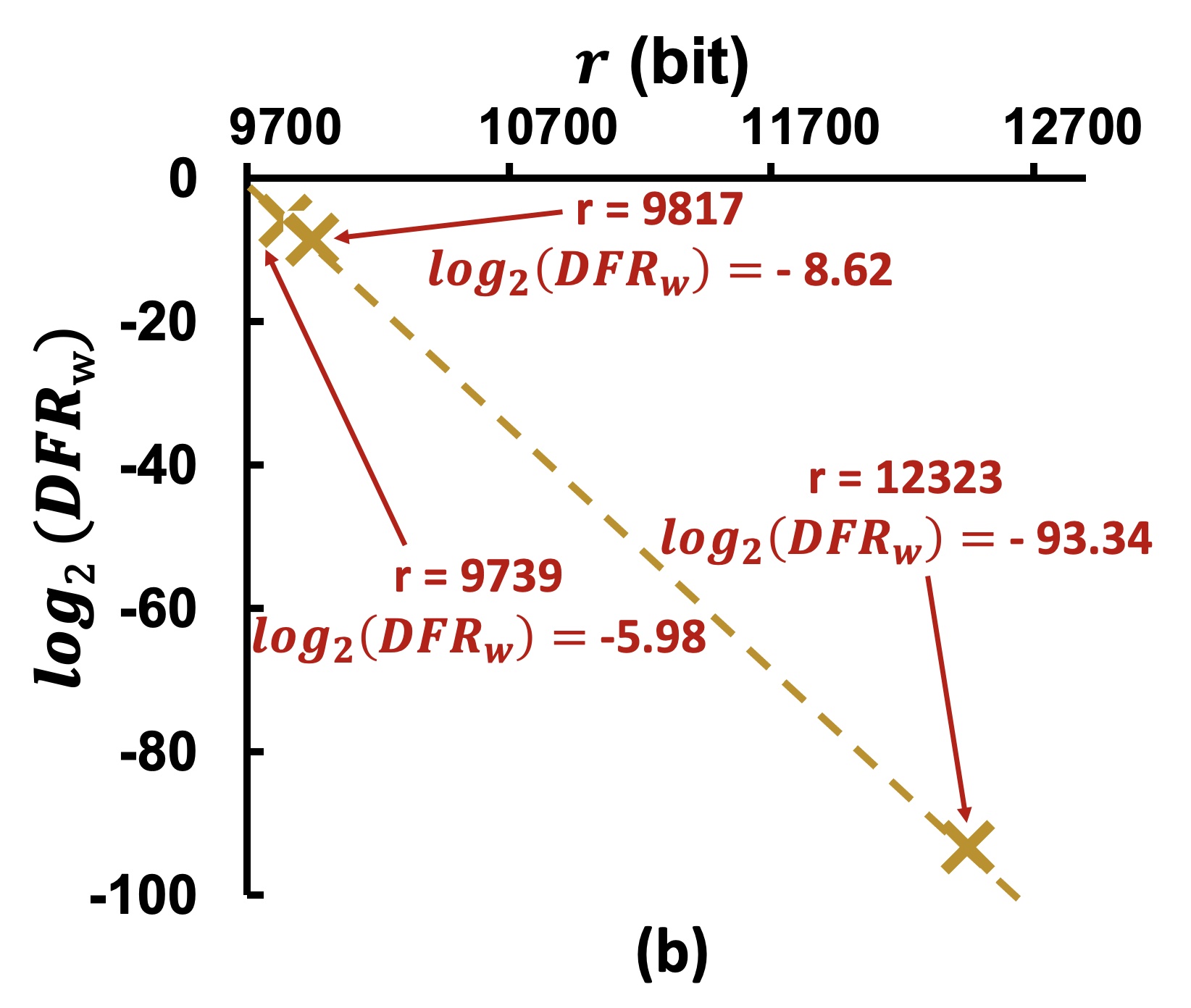}
    \includegraphics[width=0.3\textwidth, height=1.3in]{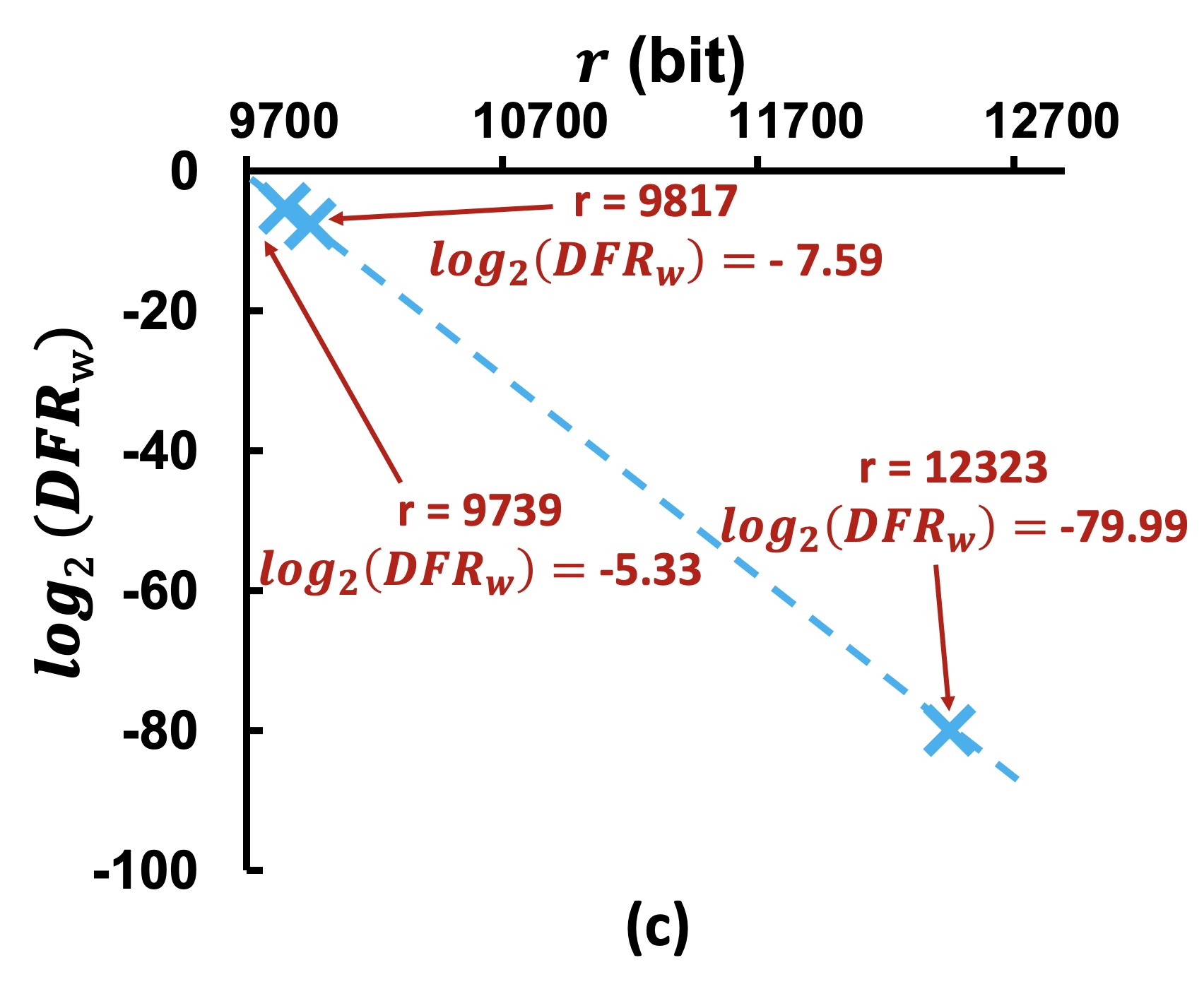}\\ 
    \includegraphics[width=0.3\textwidth, height=1.3in]{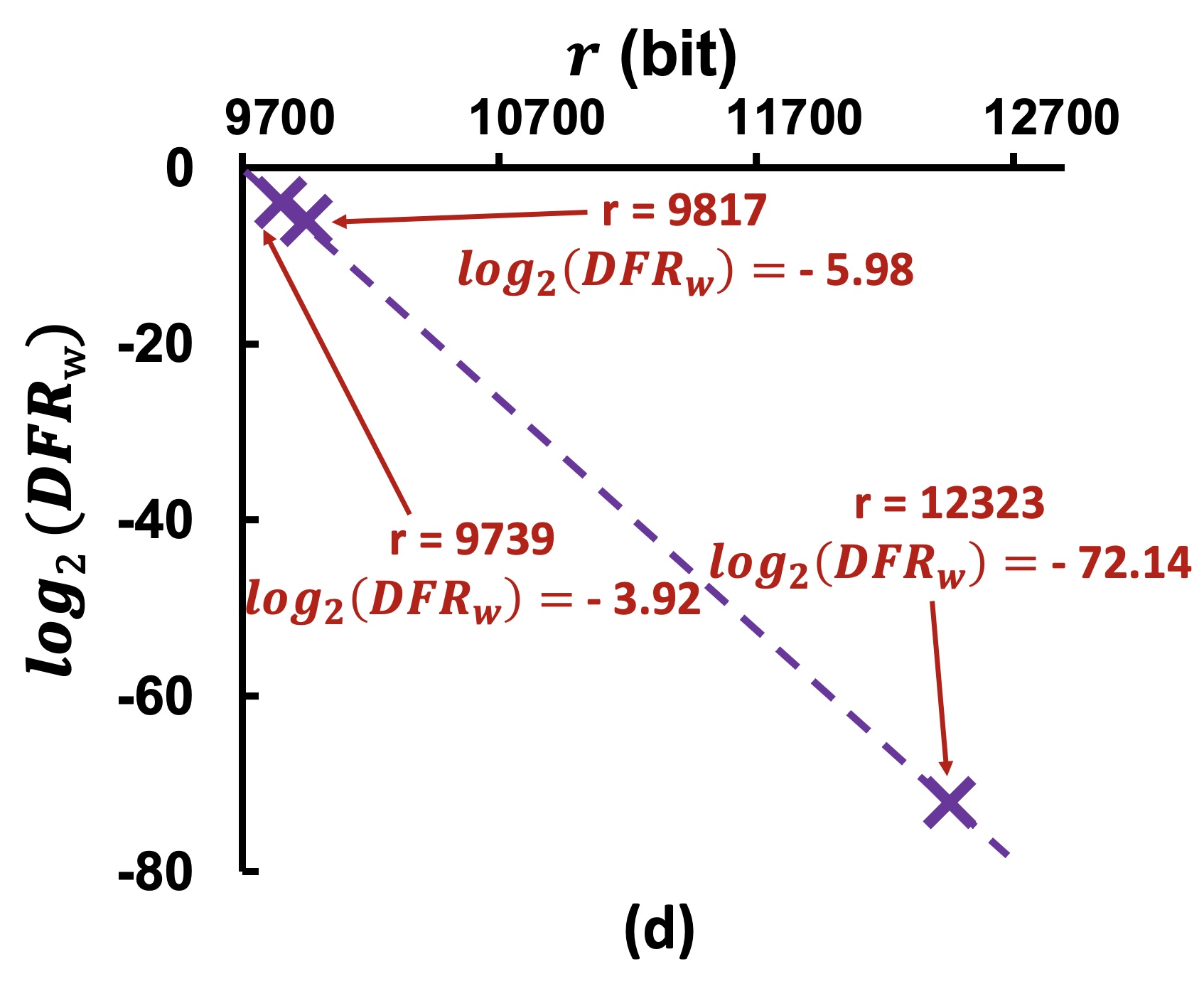}
    \includegraphics[width=0.3\textwidth, height=1.3in]{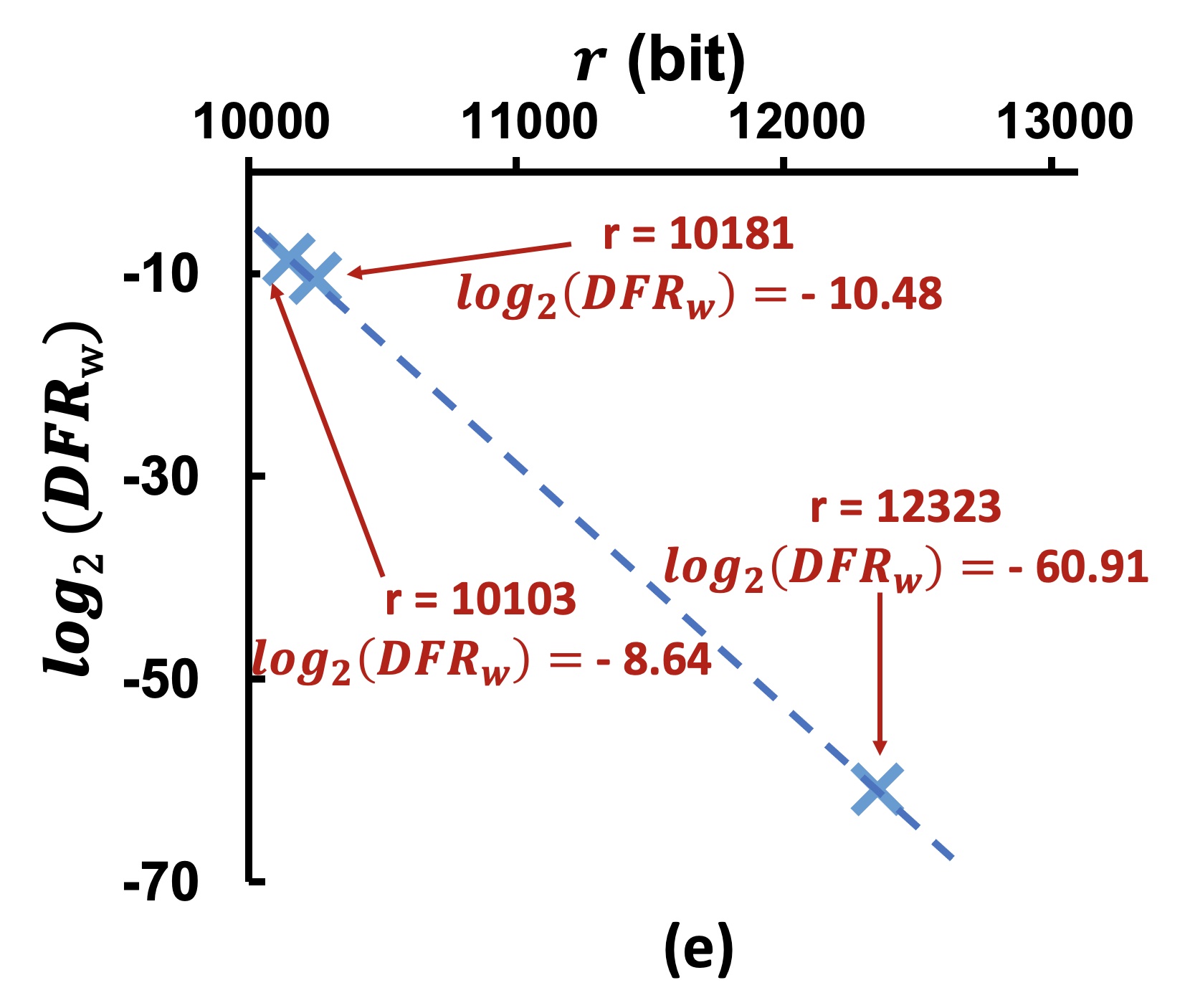}
    \includegraphics[width=0.3\textwidth, height=1.3in]{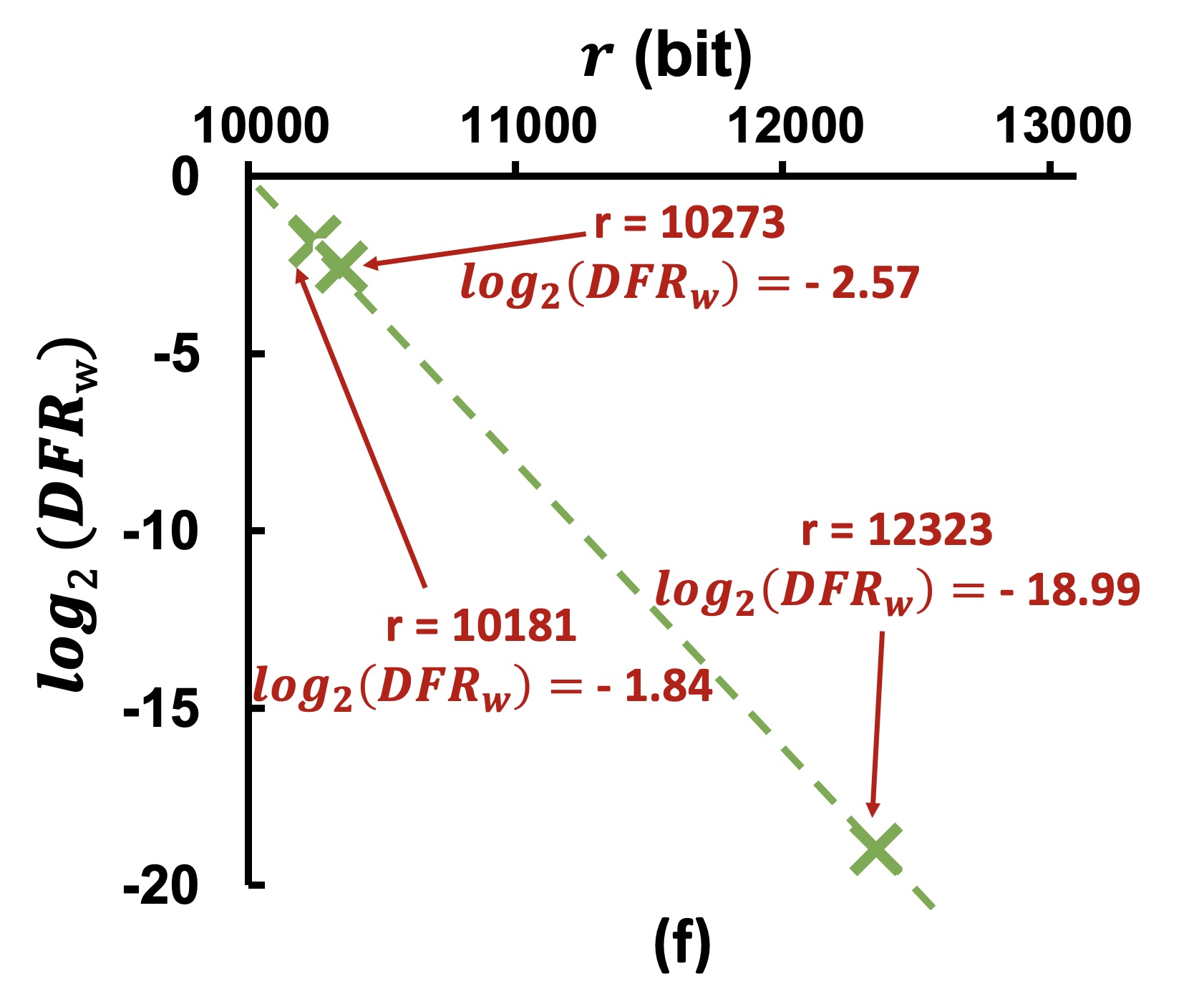}
    \caption{Result of linear extrapolation technique proposed in \cite{sendrier2020low} for estimation of DFR with varying values of $f$. (a) $f=5$, (b) $f=10$, (c) $f=15$, (d) $f=20$, (e) $f=25$, (f) $f=30$.}
    \label{fig:DFR}
\end{figure*}
\subsection{Experimental Methodology}
As mentioned before, there exists no formal mathematical model that can lead to precise computation of DFR in BF-based decoders. To circumvent this, in prior works, DFR is estimated empirically. We adopt a similar empirical approach for DFR estimation in this work, but with the emphasis on weak-keys, a feature that has been overlooked in prior works, specifically in context of BGF decoders (i.e., recommended for BIKE mechanism submitted to NIST). To conduct our analysis and visualize the impact of weak-keys upon BIKE (i.e., BGF decoder), we leveraged our Matlab implementation as detailed above in Section \ref{impl}. We started our analysis by crafting Type I weak-keys (see Section \ref{t1}) for varying values of $r$ (this is selected such that 2 is primitive modulo $r$) and by incrementing parameter $f$ from 5 to 40 in steps of 5 for each value of $r$. 

Ideally, one needs to perform an analysis on the value of $r$ that will result in the DFR corresponding to the required level of security. This is because the prior research has shown that the average DFR must be upper bound by $2^{-\lambda}$ for ensuring the $\lambda$-bit of IND-CCA security \cite{sendrier2020low}. Therefore, for 128-bit of security (i.e., minimum requirement for NIST standardization), the DFR of deployed decoder should be $2^{-128}$. In other words, $2^{128}$ ciphertexts must be generated and applied to the decoder to record a single failure, which is impracticable even on a powerful and efficient computing platform. In view of this bottleneck, prior research (see \cite{sendrier2020low}) have resorted to extrapolation techniques applied to DFR curve obtained with some small values of $r$ (as compared with that needed for DFR of $2^{-128}$, but sufficiently large to estimate the overall trend of DFR). This technique is based on the assumption (supported by empirical data) that $log(DFR_{\lambda}(r))$ is a concave and decreasing function for $DFR_{\lambda}(r)\geq 2^{-\lambda}$.  More precisely, in this approach, DFR is empirically obtained for some smaller values of $r$ (which results in relatively large DFRs that can be measured using simulation), and then, the last two points on the DFR curve are linearly extrapolated to obtain the third point that corresponds to the desired value of $r$ needed for target security level (e.g., $r=12323$ in the BIKE BGF decoder for 128-bit security). 

We adopt a similar methodology for our analysis --- i.e., for each value of $f$ (i.e., 5 to 40), we compute DFR with two relatively small values of $r$ before extrapolating them to $r = 12323$ (i.e., the same $r$ value proposed in BIKE \cite{BIKE} corresponding to DFR of $2^{-128}$). Precisely, we compute DFR at each point with at least 1000 failures, ensuring the confidence interval of 95\% \cite{sendrier2020low}. Our analysis revealed that, as per expectations, $DFR_{w}$ increases with $f$, thereby allowing us to move the tested values of $r$ from $r_1=9739$ and $r_2=9817$ (for $f=5, 10, 15$, and $20$) to $r_1=10103$ and $r_2=10181$ (for $f=25$), and $r_1=10181$ and $r_2=10273$ (for $f=30$). Moreover, for $f=35$ and $f=40$, since we expected large values for $DFR_w$, we did not need to perform the extrapolation approach, i.e., DFR was directly measured at $r=12323$. It is noteworthy that, the $r$ values used in our work are different from the $r$ values used in other works (e.g., in ~\cite{drucker2019constant}). Given that DFR is quadratic in nature (and not linear), our DFR estimations are not directly comparable to those reported in other articles. 

Once we obtain DFR for each value of $f$ (5-40), Eq.~(\ref{eq.Basic_cond}) suggests that for providing the $\lambda$-bit of IND-CCA security, the $P_w = \eta_w.DFR_w$ term must be smaller than $2^{-\lambda}$, where $\eta_w= \frac{\mathcal{|K}_w|}{\mathcal{|K|}}$. If $P_w$ is less than $2^{-128}$ (i.e., for 128 bit security), then weak-keys have no significant impact upon DFR and thus on the BIKE IND-CCA security. Otherwise (i.e., $P_w$ is greater than $2^{-128}$), the IND-CCA security of BIKE is impacted by weak-keys and thus can potentially be of concern.  

\subsection{Results}
\begin{figure*}[t!]
    \centering
    \includegraphics[width=0.3\textwidth, height=1.2in]{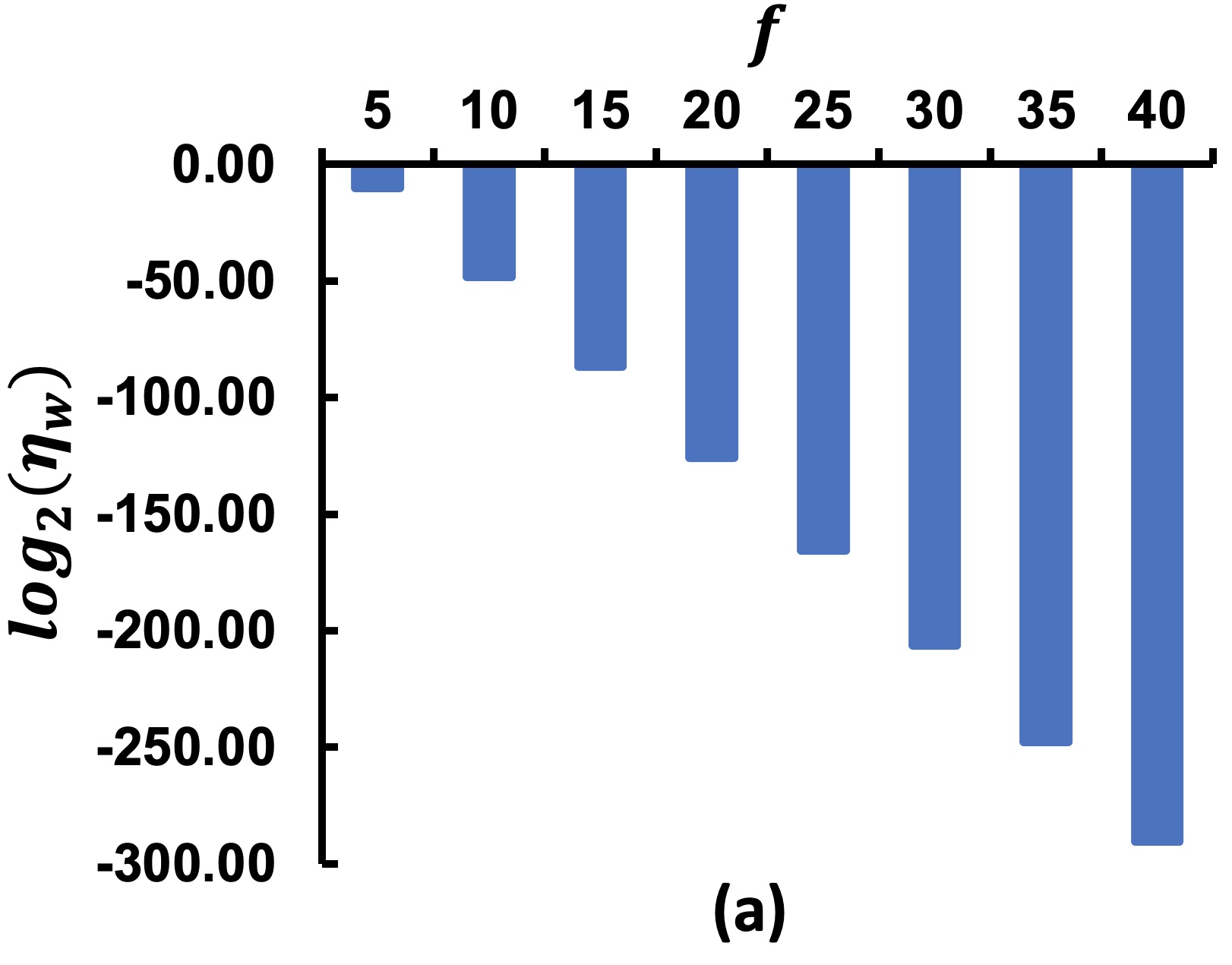}
    \includegraphics[width=0.3\textwidth, height=1.2in]{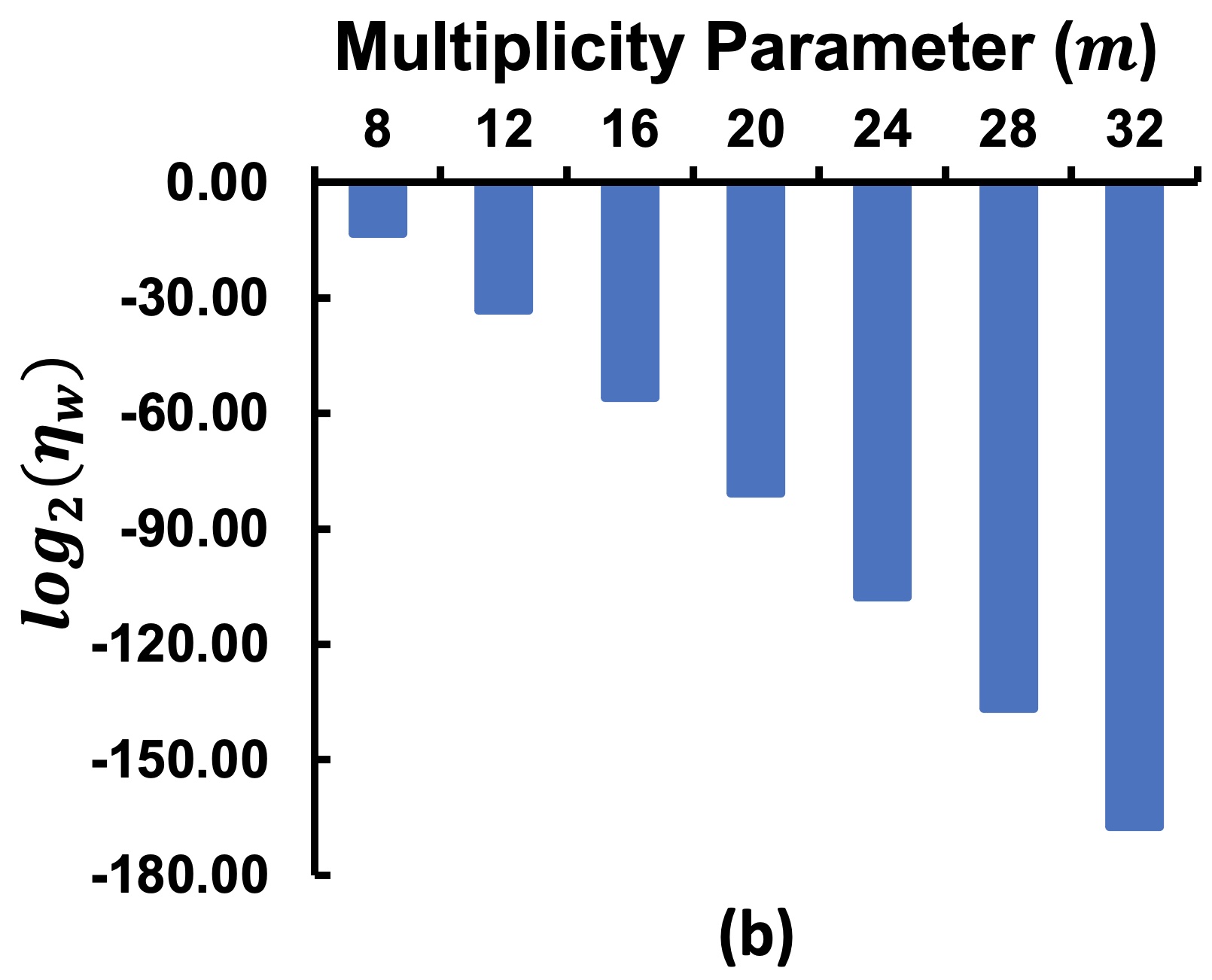}
    \includegraphics[width=0.3\textwidth, height=1.2in]{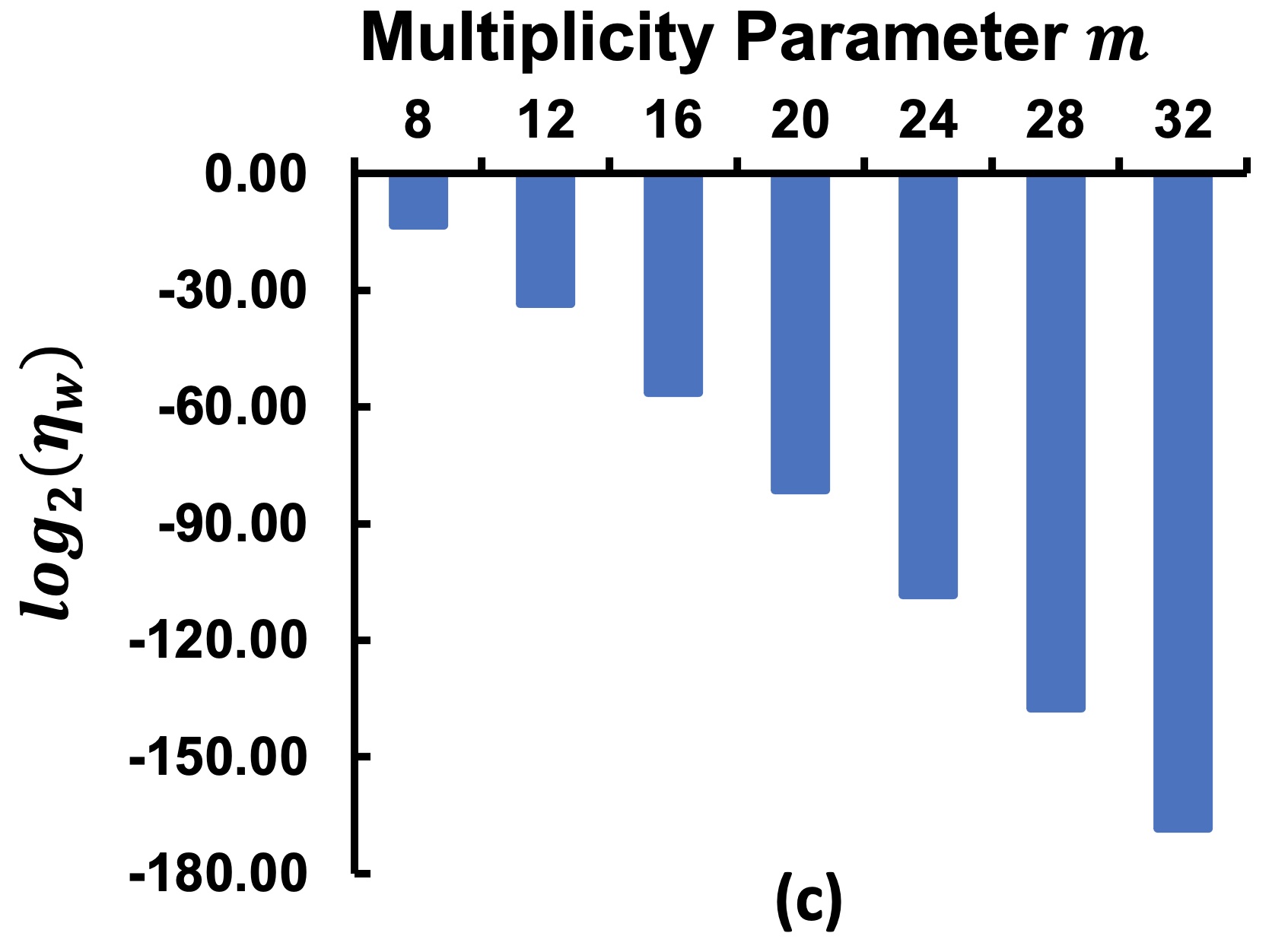} 
    \caption{ $\eta_w$ for three types of weak-keys: (a) Type 1, (b) Type 2, and (c) Type 3.}
    \label{fig:Eta}
\end{figure*}

Fig.~\ref{fig:DFR} presents the actual $DFR_w$ (i.e., $log_2(DFR_w)$) for two values of $r$ that we tested for each value of $f$ (i.e., 5 to 40) along with the results of linear extrapolation for $r = 12323$ (i.e., proposed value in BIKE \cite{BIKE} for 128-bit security). As is evident from Figure \ref{fig:DFR}, the DFR corresponding to $r = 12232$ (i.e, obtained through linear extrapolation) increases with $f$ (see $log_2(DFR_w) = -96.28$ for $f=5$ vs $log_2(DFR_w) = -18.99$ for $f= 30$). For larger values of $r$, i.e., 35 and 40, the corresponding values of $DFR_w$ are observed to be 0.8 and 1 ($log_2(DFR_w)$ will be -0.32 and 0), respectively, even when no extrapolation is performed and tested directly with $r=12323$. For gaining the insights into the IND-CCA security of BIKE mechanism in presence of weak-keys, we are interested in finding the values of term $P_w = \eta_w.DFR_w$. In accordance with our prior discussion presented in sub-sections \ref{t1} - \ref{t3} (see equations \ref{eq:t1} - \ref{eq:t3}), $\eta_w$ in this equation is dependent upon $f$ (in Type I weak-keys) and multiplicity parameter (in Type II and III weak-keys). The values of $\eta_w$ for Type I - Type III weak-keys with varying parameters (i.e., $f$ and multiplicity factor) are computed using Eqs.~\ref{eq:t1} - \ref{eq:t3}, and shown in Fig.~\ref{fig:Eta}. As expected, for large parameters $f$ and $m$, the number of weak-keys are negligible as compared with the size of the whole key space. This results in negligible values for $\eta_w$ at larger values of $f$ and $m$.

\begin{table}[!ht]
	\caption{The effect of Type I weak-keys on IND-CCA security of the BIKE scheme with parameters $(r,w,t)=(12323, 142, 134)$ and $\lambda=128$ (see Equation (\ref{eq.Basic_cond})).}
	\label{Experiments}
	\begin{center}
	\resizebox{0.42\textwidth}{!}{%
		\begin{tabular}{|c|c|c|c|}
		\hline
			\bfseries $f$ & \bfseries $log_2(\eta_w)$ & \bfseries $log_2(DFR_w)$& \bfseries $\eta_w.DFR_w$\\
            \hline
            5 & -10.225 & -96.28 & $2^{-106.50}$\\
            \hline
            $10$ & $-48.168$ & $-93.34$ & $2^{-141.51}$\\
            \hline
            15 & -86.6952 & -79.99 & $2^{-166.69}$\\
            \hline
            20 & -125.8586 & -72.14 & $2^{-198.00}$\\
            \hline
            25 & -165.7205 & -60.91 & $2^{-226.63}$\\
            \hline
            30 & -206.3566 & -18.99 & $2^{-225.35}$\\
            \hline
            35 & -247.8609 & -0.32 & $2^{-248.18}$ \\
            \hline
            40 & -290.3535 & 0 & $2^{-290.35}$\\
            \hline
		\end{tabular}%
		}
	\end{center}
\end{table}

Table \ref{Experiments} summarizes the impact of Type 1 weak-keys on IND-CCA security of the BIKE scheme.  As is evident from Table 2, for $f=5$, the term $P_w = \eta_w.DFR_w$ results in a value that is not appropriate for ensuring the 128-bit of security (i.e., minimum requirement for NIST standardization). Precisely, for $f=5$, the corresponding value of $P_w$ is $2^{-106.50}\nleq2^{-128}$, which suggests that the corresponding security level with this parameter is not adequate for NIST standardization. In fact, the contribution of Type 1 weak-keys with $f=5$ in the average DFR of the decoder is larger than the maximum allowed value. For re-affirming our observations, we repeated our experiments for $f=5$, this time with $r=10009$ which is closer to $r=12232$ and performed the extrapolation procedure with $r=9817$ and $r=10009$. The result shows that DFR (at $r=12323$) in this case increases further to -80.5 as compared with previously obtained value of -96.28. This again re-confirms that the term $P_w = \eta_w.DFR_w$ in this case is greater than required value of $2^{-128}$ (i.e, $2^{-90.73}$ in this case). It is noteworthy that, we only obtained 13 failures out of 0.3M ciphertexts in this case (vs 1000 in previous case) due to a larger value of $r$. However, our repeated experiments with decoders suggests that failure rate generally stays consistent when tested with sufficiently large numbers of ciphertexts. Also, note that, the overall contribution of Type 1 weak-keys on the average DFR is the sum of values shown in the last column of the Table 2 (as well as the values for other $f$s that we have not considered in this work, e.g., $f=6, 7, 8$, etc), which may further impact the overall average DFR negatively. However, only the first row of the Table 2 suffice to demonstrate the negative effect of weak-keys on IND-CCA security of the BIKE scheme that may need an immediate attention of the research community prior to NIST standardization. Note that, we have not repeated our experiments for Types II and III weak-keys. We expect similar results for small values of $m$ in Types II and III (e.g., $m=8$) since their dependent parameter (i.e., multiplicity) exhibits a correlation with $f$ of Type I weak-keys (see Fig.~\ref{fig:Eta}). We leave this analysis for future to have the definitive insights.    

\section{A Key-Check Algorithm}\label{Section.5}
In light of our analysis mentioned above, herein, we propose an algorithm that can potentially supplement BIKE mechanism in selecting private keys that are not weak, and thus can aid in ensuring the IND-CCA security. The proposed algorithm is based on the structure of each type of weak-keys reviewed in Section \ref{Section.3}. The detailed working of the algorithm is depicted in Algorithm 1 and each involved step is explained below. \par

Assume that $\textbf{h}=(\textbf{h}_0 \ \textbf{h}_1)$ indicates a private key generated by a user, where $\textbf{h}_i$ ($i \in\{0,1\}$) is a polynomial with the maximum degree of $r-1$. We consider each polynomial $\textbf{h}_i$ as $\textbf{h}_i=\sum_{j\in Supp(\textbf{h}_i)}x^j$, where $Supp(\textbf{h}_i)=\{p_{1}^{(i)},p_{2}^{(i)},\ldots,p_{w/2}^{(i)}\}$ is called the support of $\textbf{h}_i$ that includes the positions of non-zero coefficients in $\textbf{h}_i$ (or non-zero bits in the corresponding binary vector). The proposed key-check algorithm takes $\textbf{h}_i$ ($i \in\{0,1\}$) as the input and returns either $Weak$ or $Normal$ as the output. It first initializes the distance vector $\textbf{D}=\{D_1,D_2,\ldots\,D_{\lfloor r/2\rfloor}\}$ by assigning 0 to all its elements, i.e., $D_d=0$ ($d\in\{1,2,\ldots,\lfloor r/2\rfloor\}$). Note that, at the end of the algorithm, every element $D_d$ will indicate the multiplicity of distance $d$ ($d\in\{1,2,\ldots,\lfloor r/2\rfloor\}$) in the private key. Thus, the largest element of $\textbf{D}$ will be compared with a specific threshold (e.g., 5, 10, etc.) to decide whether it should be considered as $Weak$ or $Normal$. \par

\begin{table}
	\label{tab:freq}
	\resizebox{0.5\textwidth}{!}{%
	\begin{tabular}{l}
		\hline
		\textbf{Algorithm 1:} Key-Check Algorithm\\
		\hline
		\textbf{Input:} Multiplicity threshold $T$, $\textbf{h}_i=\sum_{j\in Supp(\textbf{h}_i)}x^j$ for $i\in\{0,1\}$ \\
		\ \ \ \ \ \ \ \ \ \ and $Supp(\textbf{h}_i)=\{p_{1}^{(i)},p_{2}^{(i)},\ldots,p_{w/2}^{(i)}\}$, $r$, and $w$\\
		\textbf{Output:} $Weak$ or $Normal$\\
		1: \textbf{for} $i=0:1$\\
		2: \ \ \ $\textbf{D}=\{D_1, D_2, \ldots, D_{\lfloor r/2\rfloor}\}=\textbf{0}$\\
		3: \ \ \ \textbf{for} $j=1:w/2$ \\
		4: \ \ \ \ \ \ \textbf{for} $k=j+1:w/2$ \\
		5: \ \ \ \ \ \ \ \ \ $temp=distance(p_j^{(i)},p_k^{(i)})$ \\
		\ \ \ \ \ \ \ \ \ \ \ \ \ \ \ \ \ \ \ \ $=min[(p_j^{(i)}-p_k^{(i)}+r \ mod \ r), (p_k^{(i)}-p_j^{(i)}+r \ mod \ r)]$\\
		6: \ \ \ \ \ \ \ \ \ $D_{temp}=D_{temp} + 1$\\
		7: \ \ \ \textbf{if} $max(\textbf{D}) > T:$\\
		8: \ \ \ \ \ \ \textbf{return} $weak$ \\
		9: \textbf{for} $j=1:w/2$\\
		10: \ \ \ \textbf{for} $k=1:w/2$\\
		11: \ \ \ \ \ \ \textbf{if} $|\textbf{h}_0\star x^{p_j^{(0)}-p_k^{(1)}}\textbf{h}_1|>T$ \\
		12: \ \ \ \ \ \ \ \ \ \textbf{return} $Weak$ \\
		13: \textbf{return} $Normal$
		\end{tabular}
		}
\end{table}

For every block $\textbf{h}_i$ ($i \in \{0,1\}$), the algorithm computes distance between the position of every non-zero coefficient (specified by $p_j^{(i)}$) and the positions of the remaining non-zero coefficients located at the right-hand side (note that the \textbf{for} loop in line 4 starts at $j+1$). Then, the multiplicity counter $D_{temp}$ associated with the computed distance $temp$ is increased by 1. This is repeated for all the $w/2$ non-zero coefficients of $\textbf{h}_i$. Finally, if the maximum element of $\textbf{D}$ is larger than the multiplicity threshold $T$, the key is identified as a weak-key.\par 

Lines 9-13 of the algorithm checks the key against the weak-key structure of Type III. In Type III, the component-wise product of block $\textbf{h}_0$ and $\sigma$-bit shifts of block $\textbf{h}_0$ must have a size larger than $T$, where $\sigma$ is the distance between the position of non-zero coefficients in $\textbf{h}_0$ (specified by $p_j^{(0)}$) and the positions of non-zero coefficients in $\textbf{h}_1$ (specified by $p_k^{(1)}$). In fact, applying a circular shift of $p_j^{(0)}-p_k^{(1)}$ bits on $\textbf{h}_1$ (by applying the $x^{p_j^{(0)}-p_k^{(1)}}$ term) will move the non-zero coefficient of $\textbf{h}_1$ located at position $p_k^{(1)}$ to position $p_j^{(0)}$. In this case, the two polynomial ($\textbf{h}_0$ and $x^{p_j^{(0)}-p_k^{(1)}}\textbf{h}_1$) will have a non-zero coefficient at the same position $p_j^{(0)}$ which may generate a high number of intersections between the corresponding columns in the parity-check matrix. Finally, if the multiplicity of intersections is less than the threshold, the key is identified as a normal key.     

\section{Conclusion}\label{Section.6}
This paper investigated the impact of weak-keys on IND-CCA security of the BIKE post-quantum key encapsulation mechanism. We first implemented the BIKE scheme with the parameters suggested in the BIKE technical specifications. Then, we performed extensive experiments to estimate the DFR of the BIKE BGF decoder for weak-keys. Our analysis suggests that the weak-keys are precarious for the IND-CCA security of the BIKE scheme and thus need immediate attention from the relevant research community prior to NIST standardization. We believe that this issue can be addressed by a potential key-check algorithm that we propose to supplement the BIKE mechanism. Theoretically, our key-check algorithm can prevent users from adopting weak private keys. The empirical analysis of the key-check algorithm and Type II and III weak-keys are left for future work to have an affirmative understanding of the effect of weak-keys on the BIKE mechanism.

\section*{Acknowledgement}
\noindent This work was partially funded by the Cyber Security Cooperative Research Centre (CSCRC).

\noindent We thank Dr. Nir Drucker (IBM Research - Haifa) for providing useful comments on the paper.

\bibliographystyle{IEEEtran}
\bibliography{TIFS}

\begin{thebibliography}{10}
\providecommand{\url}[1]{#1}
\csname url@samestyle\endcsname
\providecommand{\newblock}{\relax}
\providecommand{\bibinfo}[2]{#2}
\providecommand{\BIBentrySTDinterwordspacing}{\spaceskip=0pt\relax}
\providecommand{\BIBentryALTinterwordstretchfactor}{4}
\providecommand{\BIBentryALTinterwordspacing}{\spaceskip=\fontdimen2\font plus
\BIBentryALTinterwordstretchfactor\fontdimen3\font minus
  \fontdimen4\font\relax}
\providecommand{\BIBforeignlanguage}[2]{{%
\expandafter\ifx\csname l@#1\endcsname\relax
\typeout{** WARNING: IEEEtran.bst: No hyphenation pattern has been}%
\typeout{** loaded for the language `#1'. Using the pattern for}%
\typeout{** the default language instead.}%
\else
\language=\csname l@#1\endcsname
\fi
#2}}
\providecommand{\BIBdecl}{\relax}
\BIBdecl

\bibitem{rivest1978method}
R.~L. Rivest, A.~Shamir, and L.~Adleman, ``A method for obtaining digital
  signatures and public-key cryptosystems,'' \emph{Communications of the ACM},
  vol.~21, no.~2, pp. 120--126, 1978.

\bibitem{koblitz1987elliptic}
N.~Koblitz, ``Elliptic curve cryptosystems,'' \emph{Mathematics of
  computation}, vol.~48, no. 177, pp. 203--209, 1987.

\bibitem{shor1994algorithms}
P.~W. Shor, ``Algorithms for quantum computation: discrete logarithms and
  factoring,'' in \emph{Proceedings 35th annual symposium on foundations of
  computer science}.\hskip 1em plus 0.5em minus 0.4em\relax Ieee, 1994, pp.
  124--134.

\bibitem{CRQC}
``Quantum computing and post-quantum cryptography,''
  \url{https://bit.ly/3Huvuaq}.

\bibitem{mosca2018cybersecurity}
M.~Mosca, ``Cybersecurity in an era with quantum computers: Will we be ready?''
  \emph{IEEE Security \& Privacy}, vol.~16, no.~5, pp. 38--41, 2018.

\bibitem{barker2020getting}
W.~Barker, W.~Polk, and M.~Souppaya, ``Getting ready for post-quantum
  cryptography: Explore challenges associated with adoption and use of
  post-quantum cryptographic algorithms,'' \emph{the publications of NIST Cyber
  Security White Paper (DRAFT), CSRC. NIST. GOV}, vol.~26, 2020.

\bibitem{daniel2015initial}
A.~Daniel, B.~Lejla \emph{et~al.}, ``Initial recommendations of long-term
  secure post-quantum systems,'' \emph{PQCRYPTO. EU. Horizon}, vol. 2020, 2015.

\bibitem{fernandez2019pre}
T.~M. Fern{\'a}ndez-Caram{\'e}s, ``From pre-quantum to post-quantum iot
  security: A survey on quantum-resistant cryptosystems for the internet of
  things,'' \emph{IEEE Internet of Things Journal}, vol.~7, no.~7, pp.
  6457--6480, 2019.

\bibitem{8012331}
N.~Sendrier, ``Code-based cryptography: State of the art and perspectives,''
  \emph{IEEE Security \& Privacy}, vol.~15, no.~4, pp. 44--50, 2017.

\bibitem{mceliece1978public}
R.~J. McEliece, ``A public-key cryptosystem based on algebraic,'' \emph{Coding
  Thv}, vol. 4244, pp. 114--116, 1978.

\bibitem{misoczki2013mdpc}
R.~Misoczki, J.-P. Tillich, N.~Sendrier, and P.~S. Barreto, ``Mdpc-mceliece:
  New mceliece variants from moderate density parity-check codes,'' in
  \emph{2013 IEEE international symposium on information theory}.\hskip 1em
  plus 0.5em minus 0.4em\relax IEEE, 2013, pp. 2069--2073.

\bibitem{BIKE}
``Official web page of bike suite,'' \url{https://bikesuite.org}.

\bibitem{NISTReport}
\BIBentryALTinterwordspacing
``Status report on the second round of the nist post-quantum cryptography
  standardization process,'' \emph{National Institute of Standards and
  Technology (NIST)}, 2020. [Online]. Available:
  \url{https://doi.org/10.6028/NIST.IR.8309}
\BIBentrySTDinterwordspacing

\bibitem{AWS}
``Round 2 post-quantum tls is now supported in aws kms,''
  \url{https://aws.amazon.com/blogs/security/round-2-post-quantum-tls-is-now-supported-in-aws-kms/}.

\bibitem{gallager1962low}
R.~Gallager, ``Low-density parity-check codes,'' \emph{IRE Transactions on
  information theory}, vol.~8, no.~1, pp. 21--28, 1962.

\bibitem{guo2016key}
Q.~Guo, T.~Johansson, and P.~Stankovski, ``A key recovery attack on mdpc with
  cca security using decoding errors,'' in \emph{International conference on
  the theory and application of cryptology and information security}.\hskip 1em
  plus 0.5em minus 0.4em\relax Springer, 2016, pp. 789--815.

\bibitem{nilsson2018error}
A.~Nilsson, T.~Johansson, and P.~S. Wagner, ``Error amplification in code-based
  cryptography,'' \emph{Cryptology ePrint Archive}, 2018.

\bibitem{hofheinz2017modular}
D.~Hofheinz, K.~H{\"o}velmanns, and E.~Kiltz, ``A modular analysis of the
  fujisaki-okamoto transformation,'' in \emph{Theory of Cryptography
  Conference}.\hskip 1em plus 0.5em minus 0.4em\relax Springer, 2017, pp.
  341--371.

\bibitem{drucker2021applicability}
N.~Drucker, S.~Gueron, D.~Kostic, and E.~Persichetti, ``On the applicability of
  the fujisaki-okamoto transformation to the bike kem,'' \emph{International
  Journal of Computer Mathematics: Computer Systems Theory}, no. just-accepted,
  pp. 1--13, 2021.

\bibitem{sendrier2020low}
N.~Sendrier and V.~Vasseur, ``About low dfr for qc-mdpc decoding,'' in
  \emph{PQCrypto 2020-Post-Quantum Cryptography 11th International Conference},
  vol. 12100.\hskip 1em plus 0.5em minus 0.4em\relax Springer, 2020, pp.
  20--34.

\bibitem{sendrier2019decoding}
------, ``On the decoding failure rate of qc-mdpc bit-flipping decoders,'' in
  \emph{International Conference on Post-Quantum Cryptography}.\hskip 1em plus
  0.5em minus 0.4em\relax Springer, 2019, pp. 404--416.

\bibitem{drucker2019toolbox}
N.~Drucker and S.~Gueron, ``A toolbox for software optimization of qc-mdpc
  code-based cryptosystems,'' \emph{Journal of Cryptographic Engineering},
  vol.~9, no.~4, pp. 341--357, 2019.

\bibitem{drucker2019constant}
N.~Drucker, S.~Gueron, and D.~Kostic, ``On constant-time qc-mdpc decoding with
  negligible failure rate.'' \emph{IACR Cryptol. ePrint Arch.}, vol. 2019, p.
  1289, 2019.

\bibitem{drucker2020qc}
------, ``Qc-mdpc decoders with several shades of gray,'' in
  \emph{International Conference on Post-Quantum Cryptography}.\hskip 1em plus
  0.5em minus 0.4em\relax Springer, 2020, pp. 35--50.

\bibitem{nilsson2021weighted}
A.~Nilsson, I.~Bocharova, B.~Kudryashov, and T.~Johansson, ``A weighted bit
  flipping decoder for qc-mdpc-based cryptosystems,'' in \emph{2021 IEEE
  International Symposium on Information Theory}, 2021.

\bibitem{sendrierexistence}
N.~Sendrier and V.~Vasseur12, ``On the existence of weak keys for qc-mdpc
  decoding,'' 2020.

\bibitem{aragon2017bike}
N.~Aragon, P.~Barreto, S.~Bettaieb, L.~Bidoux, O.~Blazy, J.-C. Deneuville,
  P.~Gaborit, S.~Gueron, T.~Guneysu, C.~A. Melchor \emph{et~al.}, ``Bike: bit
  flipping key encapsulation,'' 2017.

\bibitem{zajac2014overview}
M.~Repka and P.~Zajac, ``Overview of the mceliece cryptosystem and its
  security,'' \emph{Tatra Mt. Math. Publ}, vol.~60, pp. 57--83, 2014.

\bibitem{costello2007channel}
D.~J. Costello and G.~D. Forney, ``Channel coding: The road to channel
  capacity,'' \emph{Proceedings of the IEEE}, vol.~95, no.~6, pp. 1150--1177,
  2007.

\bibitem{bonello2010low}
N.~Bonello, S.~Chen, and L.~Hanzo, ``Low-density parity-check codes and their
  rateless relatives,'' \emph{IEEE Communications Surveys \& Tutorials},
  vol.~13, no.~1, pp. 3--26, 2010.

\bibitem{fang2015survey}
Y.~Fang, G.~Bi, Y.~L. Guan, and F.~C. Lau, ``A survey on protograph ldpc codes
  and their applications,'' \emph{IEEE Communications Surveys \& Tutorials},
  vol.~17, no.~4, pp. 1989--2016, 2015.

\bibitem{berlekamp1978inherent}
E.~Berlekamp, R.~McEliece, and H.~Van~Tilborg, ``On the inherent intractability
  of certain coding problems (corresp.),'' \emph{IEEE Transactions on
  Information Theory}, vol.~24, no.~3, pp. 384--386, 1978.

\bibitem{aaronson2008limits}
S.~Aaronson, ``The limits of quantum,'' \emph{Scientific American}, vol. 298,
  no.~3, pp. 62--69, 2008.

\bibitem{baldi2014qc}
M.~Baldi, ``Qc-ldpc code-based cryptosystems,'' in \emph{QC-LDPC Code-Based
  Cryptography}.\hskip 1em plus 0.5em minus 0.4em\relax Springer, 2014, pp.
  91--117.

\bibitem{kobara2001semantically}
K.~Kobara and H.~Imai, ``Semantically secure mceliece public-key
  cryptosystems-conversions for mceliece pkc,'' in \emph{International Workshop
  on Public Key Cryptography}.\hskip 1em plus 0.5em minus 0.4em\relax Springer,
  2001, pp. 19--35.

\bibitem{bardet2016weak}
M.~Bardet, V.~Dragoi, J.-G. Luque, and A.~Otmani, ``Weak keys for the
  quasi-cyclic mdpc public key encryption scheme,'' in \emph{International
  Conference on Cryptology in Africa}.\hskip 1em plus 0.5em minus 0.4em\relax
  Springer, 2016, pp. 346--367.

\bibitem{9383383}
N.~Aydin, B.~Yildiz, and S.~Uludag, ``A class of weak keys for the qc-mdpc
  cryptosystem,'' in \emph{2020 Algebraic and Combinatorial Coding Theory
  (ACCT)}, 2020, pp. 1--4.

\bibitem{starsbars}
W.~Feller, \emph{An introduction to probability theory and its
  applications}.\hskip 1em plus 0.5em minus 0.4em\relax Wiley, 1950.

\bibitem{itoh1988fast}
T.~Itoh and S.~Tsujii, ``A fast algorithm for computing multiplicative inverses
  in gf (2m) using normal bases,'' \emph{Information and computation}, vol.~78,
  no.~3, pp. 171--177, 1988.

\bibitem{drucker2020fast}
N.~Drucker, S.~Gueron, and D.~Kostic, ``Fast polynomial inversion for post
  quantum qc-mdpc cryptography,'' in \emph{International Symposium on Cyber
  Security Cryptography and Machine Learning}.\hskip 1em plus 0.5em minus
  0.4em\relax Springer, 2020, pp. 110--127.

\end{thebibliography}
\end{document}